\documentclass{jfm}

\usepackage{graphicx}
\usepackage{natbib}
\usepackage{amsmath}
\usepackage{multirow}
\usepackage{multicol}
\usepackage{float}

\ifCUPmtlplainloaded \else
  \checkfont{eurm10}
  \iffontfound
    \IfFileExists{upmath.sty}
      {\typeout{^^JFound AMS Euler Roman fonts on the system,
                   using the 'upmath' package.^^J}%
       \usepackage{upmath}}
      {\typeout{^^JFound AMS Euler Roman fonts on the system, but you
                   dont seem to have the}%
       \typeout{'upmath' package installed. JFM.cls can take advantage
                 of these fonts,^^Jif you use 'upmath' package.^^J}%
      }
  \else
  \fi
\fi

\ifCUPmtlplainloaded \else
  \checkfont{msam10}
  \iffontfound
    \IfFileExists{amssymb.sty}
      {\typeout{^^JFound AMS Symbol fonts on the system, using the
                'amssymb' package.^^J}%
       \usepackage{amssymb}%

      }{}
  \fi
\fi

\ifCUPmtlplainloaded \else
  \IfFileExists{amsbsy.sty}
    {\typeout{^^JFound the 'amsbsy' package on the system, using it.^^J}%
     \usepackage{amsbsy}}
    {}
\fi

\newsavebox{\astrutbox}
\sbox{\astrutbox}{\rule[-5pt]{0pt}{20pt}}

\title{Study of Air-Core Vortical Flow Structure Induced by a Plughole Vortex}

\author[R. Ahmed and H. C. Lim]%
{ R\ls A\ls Y\ls H\ls A\ls N\ns A\ls H\ls M\ls E\ls D$^1$\ns \and H\ls E\ls E\ls C\ls H\ls A\ls N\ls G\ns L\ls I\ls M$^1$\thanks{Author to whom correspondence should be addressed: hclim@pusan.ac.kr}%
}

\affiliation{School of Mechanical Engineering, Pusan National University, 2, Busandaehak-ro 63beon-gil, Geumjeong-gu, Busan, 46241, Rep. of KOREA\\}

\pubyear{2020}
\volume{823}
\pagerange{787--818}
\date{13 November 2015; revised 15 January 2017; accepted 15 May 2017}
\begin{document}

\maketitle


\begin{abstract}
This paper describes a study of the generation of a plughole vortex and its consequences in a drainpipe during drainage of water from a stationary rectangular tank. The critical and minimum depths of water above the inlet of the drainpipe, where a surface dip starts to develop for drainpipes of various diameters, were examined parametrically. This study explored the following naturally occurring phenomena arising from a plughole vortex. (i) A plughole vortex initially causes a surface dip to develop towards the inlet of the drainpipe and as the surface dip approaches the inlet of the drainpipe it creates a droplet-shaped air bubble. (ii) A unique bubble transformation, i.e., from a droplet-shaped to a donut-shaped bubble ring, occurs just after the separation of the droplet-shaped air bubble from the surface dip. (iii) The donut-shaped bubble ring flows with the drain water and initially causes bubbly flow in the drainpipe. (iv) As the water head above the inlet of the drainpipe decreases, the droplet-shaped bubble size increases, and consequently, the bubble ring size increases and causes slug flow in the drainpipe. (v) As the slugs combine, the flow of the draining water eventually becomes annular flow in the drainpipe. Sounds, such as that of instantaneous fizz and bubble sink draining, were observed to be produced as a result of the bubble formation process. Temporal changes in the shape and size of the air bubbles were studied. Within the range of 0.45 to 0.6, the ratio of the bubble diameter to the bubble length was found to be linearly proportional to the ratio of the water depth to the diameter of the drainpipe. Several drainage cases were simulated numerically to observe the physics of these naturally occurring phenomena. The shapes and sizes of the vortices induced by plugholes have been visualised and analysed using the vortex core method.
\end{abstract}

\begin{keywords}
Authors should not enter keywords on the manuscript, as these must be chosen by the author during the online submission process and will then be added during the typesetting process (see http://journals.cambridge.org/data/\linebreak[3]relatedlink/jfm-\linebreak[3]keywords.pdf for the full list)
\end{keywords}

\section{Introduction}\label{sec:introduction}
A plughole or bathtub vortex, which results from an ineluctable process that occurs when water is drained from a stationary or rotating tank, induces a dip on the free surface of water, with or without a visible whirling motion in the water. The vortex flow can be readily observed in nature and is similar to other types of vortices, such as a pump intake vortex, tornado vortex, and gas core vortex in the primary cooling system of a conventional thermal nuclear reactor. Naturally, the formation of a dip on the free surface causes air to enter the drainpipe, which in most cases is undesirable. A plughole vortex can be observed not only during draining of water but also in some other situations, such as draining of molten metal in the process of discharging a tundish (an open container with one or more holes in the bottom) \citep{Park&Sohn2011}, draining of liquid propellant from a tank, and scavenging of lubrication oil from an aero-engine bearing chamber. Numerous theoretical and experimental studies have been conducted to determine the direction of rotation of a plughole vortex and the influences of various fluid properties and circumstances.\par

The direction of rotation of localised whirling during draining usually depends upon some random factors such as the residual motion in the liquid, air flow above the liquid surface, boundary conditions, temperature inhomogeneity, and the Coriolis force. The Coriolis force is negligible compared to the other factors mentioned, but when draining of liquid from a tank begins under steady-state conditions, i.e., free from any external forces, the formation of a whirlpool is solely due to the Coriolis force influenced by the Earth's rotation \citep{Lugt1983}. Under the influence of the Coriolis force, the direction of rotation of the vortex has been observed to be counter-clockwise in the Northern Hemisphere (see \citealt{Turmlitz1908, Shapiro1962, Binnie1964}), and clockwise in the Southern Hemisphere \citep{Trefethenetal1965}. The present study was conducted in South Korea, which is located in the Northern Hemisphere, and the direction of rotation observed while draining under steady-state conditions was counter-clockwise, but the rotation was negligibly small.\par

The effect of the drain flow rate on the free surface shape was studied theoretically by \cite{Miles1998}, \cite{Saad&Oliver1964}, and \cite{Bhuta&Koval1964}. In particular, Saad \& Oliver, and Bhuta \& Koval showed that draining introduces oscillations in an initially stationary free surface, whereas Miles showed that initial oscillations on a free surface are damped by the drainage of the liquid. \cite{Lubin&Springer1967} (hereinafter, LS) attempted to identify the conditions under which a dip on the liquid surface approaches the bottom of the container and enters the drainpipe. They also observed the plughole drain flow in a stationary cylindrical tank with flush-mounted drainpipes at the bottom of the tank, and found that during liquid draining a dip develops on the free surface when the fluid level reaches a certain height above the bottom of the tank, which was termed the 'critical height'. The dip formation and its development in a drainpipe were observed by the naked eye to be so rapid that it was termed as 'rapid (i.e., almost instantaneous) dip development'. In addition, the authors worked with various immiscible liquid-liquid and liquid-air combinations to examine the effect of density variation on the critical height. Based on the results of a theoretical analysis conducted using Bernoulli's equation, they proposed an empirical correlation to find the critical height for the fluid layers with respect to the density variation. \par

One of the leading theoretical studies of vertical flow above drain holes was the work of \cite{Lundgren1985}. He tried to solve the problem of modelling slow draining of an incompressible fluid from a rotating cylindrical vessel. In his theoretical model, a liquid confined between two infinite flat plates separated by a finite distance was considered, and both the plates and liquid rotated at a certain angular velocity. The mathematical model included the transient effects, internal viscosity, wall friction of the boundary region, and finite cylindrical boundaries. \par

Notable success in numerical modelling of the free surface profile was achieved by \cite{Zhou&Graebel1990}, who introduced the boundary-integral method (BIM) for a two-dimensional (2-D) axisymmetric case. This case was successfully modelled using a nonlinear BIM scheme in which the surface tension and viscosity were neglected and the surface position, pressure, and velocity values at any time interval during draining were identified. The model effectively described the time-dependent free-surface profile for various ranges of Froude numbers. \cite{Tyvand1992} analysed a potential theory concerning the free-surface flow with a single sink and found a critical sink strength for the formation of a surface dip whose Froude number was equal to $1/3$. \cite{Forbes&Hocking1995, Forbes&Hocking2007, Forbes&Hocking2010}, \cite{Stokesetal2005, Stokesetal2008, Stokesetal2012}, and \cite{Farrow&Hocking2006} studied withdrawal of fluid through a sink both analytically and numerically for a variety of cases, with a focus on the free surface profile and examined the effects of several critical parameters, such as the surface tension, Froude number, and non-dimensional sink radius, on the free surface profile. They also proposed a linearized theory for the extraction rate and a novel numerical solution technique based on a Fourier series.\par

In an effort to examine the issue of bubble formation instability in a drainpipe,\cite{Andersenetal2003} conducted both experimental and theoretical modelling of a bathtub vortex generated by the draining of a rotating cylindrical container through a small drain hole. They obtained a flow structure similar to that obtained by \cite{Lundgren1985}, although their models included surface tension and Ekman upwelling. In addition, bubble generation was observed at high rotation rates when the tip of the needle-like surface depression became unstable. \cite{Stepanyants&Yeoh2008} used a modified Lundgren model to describe a stationary bathtub vortex in a viscous liquid with a free surface. They proposed three different drainage regimes: subcritical, when the dent of the whirlpool is less than the fluid depth; critical, when the tip of the whirlpool dent touches the outlet orifice; and supercritical, when the surface vortex entrains air into the intake pipe. In addition, they observed the effect of a dimensionless Kolf number on the dimensionless liquid depth for a fixed drain flow rate. By adjusting the Kolf number for a fixed drain flow rate, they attempted to determine the subcritical and near-critical water depths. \par

The VOF (Volume Of Fluid) method has been used for surface tracking in multiphase immiscible viscous flows and free surface flows. Some notable research efforts in this area have been those by \cite{Gueyffieretal1999} and \cite{Gopala&Wachem2008}. They used a linear model to calculate the interface in each cell and computed the momentum balance and surface tension using explicit finite differences on a cubic grid and the continuous surface force method, respectively. \cite{Robinsonetal2010} used the VOF method to conduct a computational analysis of plughole vortices induced during draining of stationary vessels. They examined several combinations of tanks, drainpipes, and fluids to assess the potential use of computational fluid dynamics to describe the induced vortices. Their results were compared with the results of LS, and some deviations were observed. They described their computational approach in detail, along with their reasons for choosing various models and parameters. \par

Although significant progress has been achieved in characterizing plughole vortices and identifying some of their critical parameters, the consequences of a plughole vortex in a drainpipe have not yet been observed or analysed. Some preliminary results were reported by \cite{Ahmed&Lim2015} but were not sufficient to yield findings concerning the physics of a plughole vortex. Therefore, in this study, we sought to observe the generation and aftermath of a plughole vortex and the effects of a plughole vortex on the drain flow rate. In the generation process, the critical height of water was measured using various techniques and compared with previous experimental observations. The formation of a unique droplet-shaped bubble and its transformation from a droplet-shaped to a donut-shaped bubble ring due to the plughole vortex were observed. A computational investigation of water draining from a stationary rectangular tank and analysis of the bubble formation process and its consequences in the drainpipe were performed. Because this draining process occurs as a result of gravity and the drain flow rate remains variable, the analyses performed in the current study represented the phenomena of interest more realistically than those performed in previous studies.\par

The remainder of this paper is organised as follows. Section 2 outlines the basic configuration of the experimental setup and the experimental procedure. Section 3 describes the governing equations and computational methods used. Sections 4 and 5 compare and analyse the experimental and computational results. Section 6 summarises our major conclusions.  \par

\section{Experimental setup}\label{sec:Setup}

\begin{figure}
\centering
\includegraphics[angle=0, trim=0 0 0 0, width=1.0\textwidth]{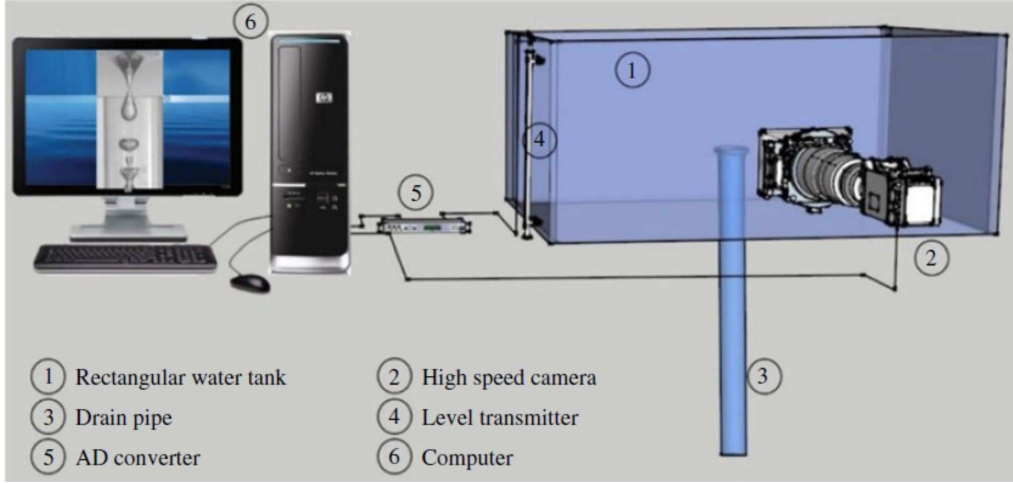}
\caption[]{Schematic diagram of experimental setup.} \label{fig:Setup}
\end{figure}

Draining experiments were carried out to observe natural draining of a liquid owing to gravity under stationary and steady-state conditions. Figure~\ref{fig:Setup} shows a schematic diagram of the experimental setup for plughole draining. A rectangular tank with dimensions of 0.6 m $\times$ 0.5 m $\times$ 0.4 m, made up of 15-mm-thick transparent acrylic sheets, was used as a stationary tank. A hole was made at the bottom centre of the tank to accommodate drainpipes of various diameters. Seven different transparent acrylic drainpipes with internal diameters of 12, 19.5, 26.2, 31.9, 39.3, 44.0, and 49.8 mm, each of which were 400-mm in length, were used to achieve different flow rates. To permit a clear view of the plughole vortex, the inlet of the drainpipe was raised 66 mm above the bottom of the tank, and the pipe outlet was blocked with a rubber pipe cap before draining. To prevent formation of an initial dent on the free surface, the tank was filled with water at some salient heights of 50 to 150 mm above the inlet to the drainpipe, depending on the diameter of the drainpipe. Note here that the initial depth of water was set to be at least 3 times greater than the drainpipe diameter; otherwise, it might have an influence on the plughole vortex formation. Initial steady-state condition of the water was achieved by allowing the water to settle for at least 45 minutes before each experiment and was confirmed by adding a colour dye to the tank. \par

A water level gauge (capacitive level transmitter, KENEK CHT6-20E), equipped with a National Instrument (NI) USB-9215 high-speed data acquisition module (DAQ) connected through a universal serial bus (USB) interface to a personal computer, was placed close to the wall inside the tank to measure the water depth during draining, and the water level was measured at a sampling rate of 50 Hz. The analysis and control software was implemented on the NI LabVIEW platform. The resolution of the data acquisition board was 16-bit, and it was capable of achieving a single-channel speed of 20-kS/s. The board used an NI signal streaming process for sustained high-speed data streams over a USB port. A scale was levelled on the inside wall of the tank to observe the water level visually. While draining, the plughole vortex influenced the surface dip to approach the inlet of drainpipe within milliseconds and transported air bubbles into the drainpipe. A high-speed camera was placed outside the tank and used to capture the air bubble generation stages and transportation into the drainpipe at a frame rate of $@$ 7,500 fps. The sound intensity level of the vortex was measured using a portable sound analyser placed within 200 mm of the source in air. A stopwatch was also used to measure the timing of each event.\par

\section{Computational methodology}\label{sec:CFD-method}

\subsection{Governing equations}
The governing equations of fluid motion are based on the Navier-Stokes equations, which in this study were formulated using the CFD commercial software ANSYS FLUENT$^{\textregistered}$ and spatially discretised using second-order finite volume methods. Numerical simulations were carried out assuming that there were no other thermal sources or stratification, and a pressure-based solver was used for this computational work, considering that the flow was mostly low-speed incompressible flow. The water flowing out through the plughole was a type of free-draining flow assisted by gravitational force, and all solutions of the numerical calculation were transient.\par

A multiphase VOF model developed by \cite{Hirt&Nichols1981} was selected for the solution of two immiscible fluid layers (i.e., water and air) sharing the same boundary and bubbly flow in the drainpipe. This VOF model was regarded as better suited for use in simulating two-phase drain flow, than a mixture model or the Eulerian model, \citep{Fluent2013}. Tracking of the interface between the fluids, air and water, was accomplished through solution of a continuity equation for the volume fraction of air or water. For the water, the volume fraction equation can be written as follows:

\begin{equation} \label{eq1}
\frac{1}{\rho_{w}} \left[ \frac{\partial}{\partial t} \left( \alpha_{w} \rho_{w} \right) + \nabla \cdot \left( \alpha_{w} \rho_{w} \vec{v}_w \right) = S_{\alpha_w} + \Sigma^{n}_{a = 1} \left( \dot{m}_{a w} - \dot{m}_{w a} \right) \right]
\end{equation} 
  
\noindent where $\alpha_w$ is the volume fraction of water, $\rho_w$ is the density of water, $\dot{m}_{w \alpha}$ is the mass transfer from water to air, $\dot{m}_{\alpha w}$ is the mass transfer from air to water, and $S_{\alpha_w}$ is the mass source term which is zero in this case. As there is no mass transfer between air and water, the right-hand side of equation \ref{eq1} is equal to zero. \par

The calculation of the drain flow requires a time-dependent scheme because the initially filled tank will empty out after a certain period of time. Therefore, an explicit time discretisation scheme was used, along with a geometric reconstruction scheme, to achieve a more accurate interface shape. As the case studied was time-dependent, these explicit and geometric reconstruction schemes matched perfectly. In the explicit scheme, standard finite-difference interpolation schemes were applied to the volume fraction values, which were computed beforehand. Equation \ref{eq1} can be discretised in time to yield the following: \par

\begin{equation} \label{eq2}
\frac{\alpha^{n+1}_{w} \rho^{n+1}_{w} - \alpha^{n}_{w} \rho^{n}_{w}}{\Delta t} V + \Sigma_f \left( \rho_w U^n_f \alpha^n_{w,f} \right) = 0 
\end{equation} 

\noindent where $n+1$ is the index for the new time step, $n$ is the index for the previous time step, $\alpha_{w,f}$ is the face value of the water volume fraction computed from the second-order upwind scheme, $V$ is the volume of the cell, and $U_f$ is the volume flux through the face based on a normal velocity. Convergence using a segregated algorithm such as the pressure-based SIMPLE (Semi-Implicit Method for Pressure Linked Equations) scheme proposed by \cite{Patankar&Spalding1972} becomes more robust when the partial equilibrium of the pressure gradient and the body forces are taken into account. To make the solution more robust, an implicit body force formation was used.\par

In transient simulation with VOF (see \citealt{Hirt&Nichols1981}), the time step used for the transport equation is not the same as the time step used for the volume fraction calculation. The time step for the VOF was automatically calculated based on the input of the maximum Courant number allowed near the free surface. The Courant number, which is dimensionless, is the ratio of the time step in the VOF calculation to the characteristic time of transit of a fluid element across a control volume. The Courant number, $C$, can be represented as follows: \par

\begin{equation} \label{eq3}
C= \frac{\Delta t}{\Delta t_{char}}
\end{equation} 

\noindent where $\Delta t$ is the time step size used in the VOF calculation and $\Delta t_{char}$ is the characteristic time. \par

In the VOF calculation, the ratio of the cell volume to the net outgoing volume flux, i.e., volume flow rate from each cell, represents the time required for the fluid to drain from the cell. The smallest time is called the characteristic time: \par

\begin{equation} \label{eq4}
\Delta t_{char} = min \left[ \Sigma \frac{\rm cell~volume}{\rm volume~flux} \right] = min \left( \frac{\Delta x_{cell}}{v_{fluid}} \right)
\end{equation}

The maximum allowable Courant Number for volume fraction calculation was set to 0.25 so that the time step for VOF calculation would be at most one-fourth the minimum transit time for any cell near the interface.\par

In addition, the global Courant number ($CFL_{global}$) was observed after each time step calculation, and an effort was made to maintain this number below 1.0 by changing the time step of the transport equation ($\Delta t_{global}$). The global Courant number, which is also dimensionless, is given by the ratio of the time step in the transport equation ($\Delta t_{global}$) to the characteristic time ($\Delta t_{char}$): \par

\begin{equation} \label{eq5}
CFL_{global} = \frac{\Delta t_{global}}{\Delta t_{char}}
\end{equation}

The $\Delta t_{global}$ values used in the simulation ranged from $10^{-4}$ to $10^{-6}$, depending on the $CFL_{global}$ number. \par

Realistic fluid viscosities of $1.7894 \times 10^{-5}  kg/ms$ and 0.001003 $kg/ms$ were used for air and water, respectively, in the simulation. In addition, to select an appropriate viscous turbulent model, the Reynolds number was calculated for flowing water through the drainpipe, as our major concern was to observe the flow pattern in the drainpipe. The calculated Reynolds number was in range of 10,000 to 150,000, which could be considered to be fully turbulent flow. Therefore, the well-known realisable k-$\varepsilon$ model proposed by \cite{Shihetal1995} was used in this analysis, with a standard wall function as the near-wall treatment scheme. The realisable k-$\varepsilon$ model, which is an improvement over the standard k-$\varepsilon$ model, generates an alternate formulation of the turbulent viscosity and a modified transport equation for the dissipation rate, $\varepsilon$. The new features of the realisable k-$\varepsilon$ model yields substantial improvement in flow features involving a strong streamline curvature, vortices, and rotation (\citealt{Kimetal1999}). \par

The two fluids, water and air, share a common interface and thus interact with the balance of the surface tension. In the simulation, the surface tension effect was modelled using the continuum surface stress method (\citealt{Lafaurieetal1994} and \citealt{Popinet&Zaleski1999}), and the surface tension coefficient was assumed to be a constant 0.072 $N/m$. \par
 
For the purpose of spatial discretisation, the least squares cell-based scheme was used for the gradient, as it is less expensive to compute the least squares gradient than the node-based gradient. The pressure discretisation was obtained using a pressure staggering method. This method is similar to the staggered grid scheme used in a structured mesh and is highly redundant when used for an unstructured mesh. A second-order upwind scheme was used for the discretisation of momentum, turbulent kinetic energy, and the turbulent dissipation rate. Under relaxation factors for the density, the body forces and turbulent viscosity were 1.0, the turbulent kinetic energy and turbulent dissipation rate were set to 0.8, and the momentum and pressure were set to 0.7 and 0.3, respectively. The residual convergence criteria were set to $10^{-7}$ for all components of velocity, continuity, k, and $\varepsilon$. Although all of the components of velocity, k, and $\varepsilon$ converged in each time step, the continuity equation did not converge in some cases. The residual was limited to $10^{-4}$. \par

\subsection{Computational grid and boundary conditions}
\begin{figure}
\centering
\includegraphics[angle=0, trim=0 0 0 0, width=1.0\textwidth]{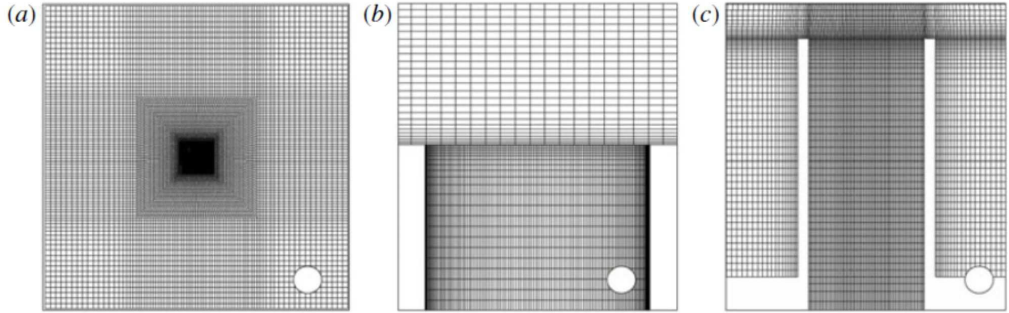}
\caption[]{Mesh structure of the draining tank: (a) plan view, (b) flush-mounted pipe, and (c) raised pipe.} \label{fig:Mesh}
\end{figure}

A draining case in a computational domain was selected for the purpose of comparing and validating the results from the real-scale experiment, whereas the other cases were selected depending on the particular interest. Figure~\ref{fig:Mesh} shows a sample mesh structure for a rectangular tank and various pipe configurations. Details of the shape and size cases are described later in this paper. A three-dimensional (3-D) computational grid was generated for test tank cases with both rectangular and circular shapes. A multi-block approach was used to mesh the tank volume, and an O-grid was used to mesh the pipe volume. In one case, the inlet of the drainpipe was raised 66 mm above the bottom of the tank, and in the other cases, to avoid mesh difficulties during the computational analysis, the inlet of the drainpipe was mounted flush with the bottom of the tank. However, in the experimental setup the inlet of the drainpipe was raised above the bottom of the tank, which had a clearly visual impact, compared to the flush-mounted drainpipe. The meshes consisted of approximately 1.7 to 2.6 million hexahedral cells clustered near the pipe inlet, maintaining the proper $y^+$ for the simulation. \par

\begin{figure}
\centering
\includegraphics[angle=0, trim=0 0 0 0, width=0.6\textwidth]{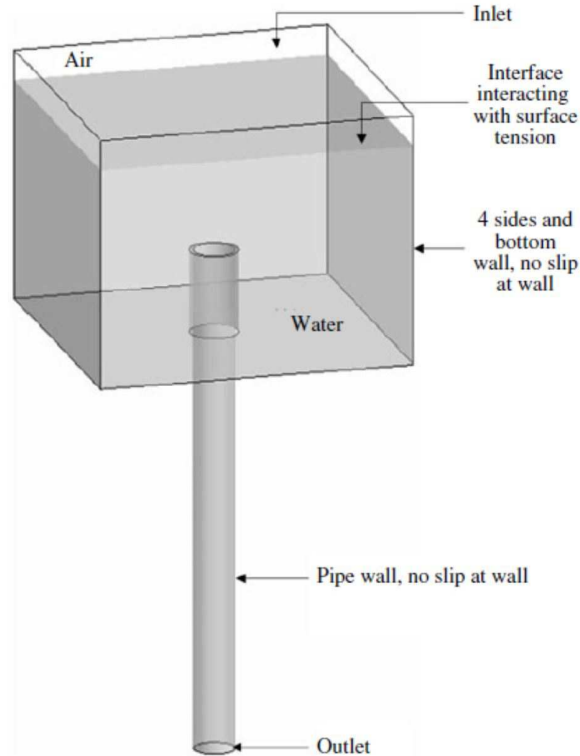}
\caption[]{Schematic of tank showing the boundary conditions.} \label{fig:BCs}
\end{figure}

Figure~\ref{fig:BCs} shows a schematic diagram of the water tank with the drainpipe that illustrates the boundary conditions used in the simulation. The boundary conditions for the inlet and outlet were set to atmospheric pressure at the top of the tank and at the bottom of the pipe, and a no slip-condition of smooth glass was imposed for all of the wall surfaces. The pipe volume and a portion of the tank volume were patched with water, assuming that part of the top volume of the tank was filled with air and the rest of the volume was filled with water. A constant surface tension was used at the interfacial boundary between the water and air.  \par

\section{Experimental results and discussion}

This study was one of the first in which the generation and consequences of plughole vortex and bubble dynamics were observed to determine what happens when a surface dip enters a drainpipe. The experimental observations are organised according to the following subject:
\begin{itemize}
\item Plughole vortex generation process: This subject pertains to the generation of a plughole vortex and the initiation of the surface dip development during draining of water from a stationary tank. Three different techniques for measuring the critical height are described (Section 4.1).
\item Bubble formation process: This subject pertains to the consequences of the plughole vortex, i.e., what happens when a surface dip enters a drainpipe (Sections 4.2-4.4). 
\end{itemize}

\subsection{Critical height}

Three different techniques were used to estimate the critical height: measurement of drain flow rate, observation of the water depth (i.e., flow visualisation), and use of the LS correlation. To measure the drain flow rate, a level gauge was placed inside the water tank to measure the temporal variation of the water level. The water depth was visualised using a high speed camera to measure the free surface position precisely as the formation of dip was not identifiable by level gauges. The LS approach was used to estimate the water depth based on an empirical correlation.

\subsubsection{Measurement of critical height using drain flow rate}

All of the experiments were conducted for a series of diameters using seven different drainpipes, and for each drainpipe test, a complete drain was conducted repeatedly (at least four to five times) with some initial water depth, which was chosen to be at least 3 times greater than the drainpipe diameter. In each experiment, the draining was started suddenly by removing the pipe cap from the bottom of the drainpipe. During the draining process, the water depth above the inlet of the drainpipe decreased gradually with time. This temporal variation of the water depth was measured experimentally and calculated theoretically. For the theoretical calculation, the depth at time $t$ was calculated using the following transient Bernoulli equation (\citealt{Finnemore&Franzini2001}):

\begin{equation} \label{eq6}
H_f = \left( \sqrt{H_i} - \frac{t A_p C_d}{A_t} \sqrt{\frac{g}{2}} \right)^2
\end{equation}

\noindent where $A_t$, $A_p$, $C_d$, $H_i$, and $H_f$  are the fixed plan area in the rectangular tank, concentric drainpipe area, discharge coefficient, initial depth at time $t = 0$, and final depth at time $t$, respectively. In the theoretical calculation, the discharge coefficient $C_d$, obtained from the curve fitting, was found to increase with the pipe diameter over a range of 0.725 to 0.78. \par

\begin{figure}
\centering
\includegraphics[angle=0, trim=0 0 0 0, width=0.8\textwidth]{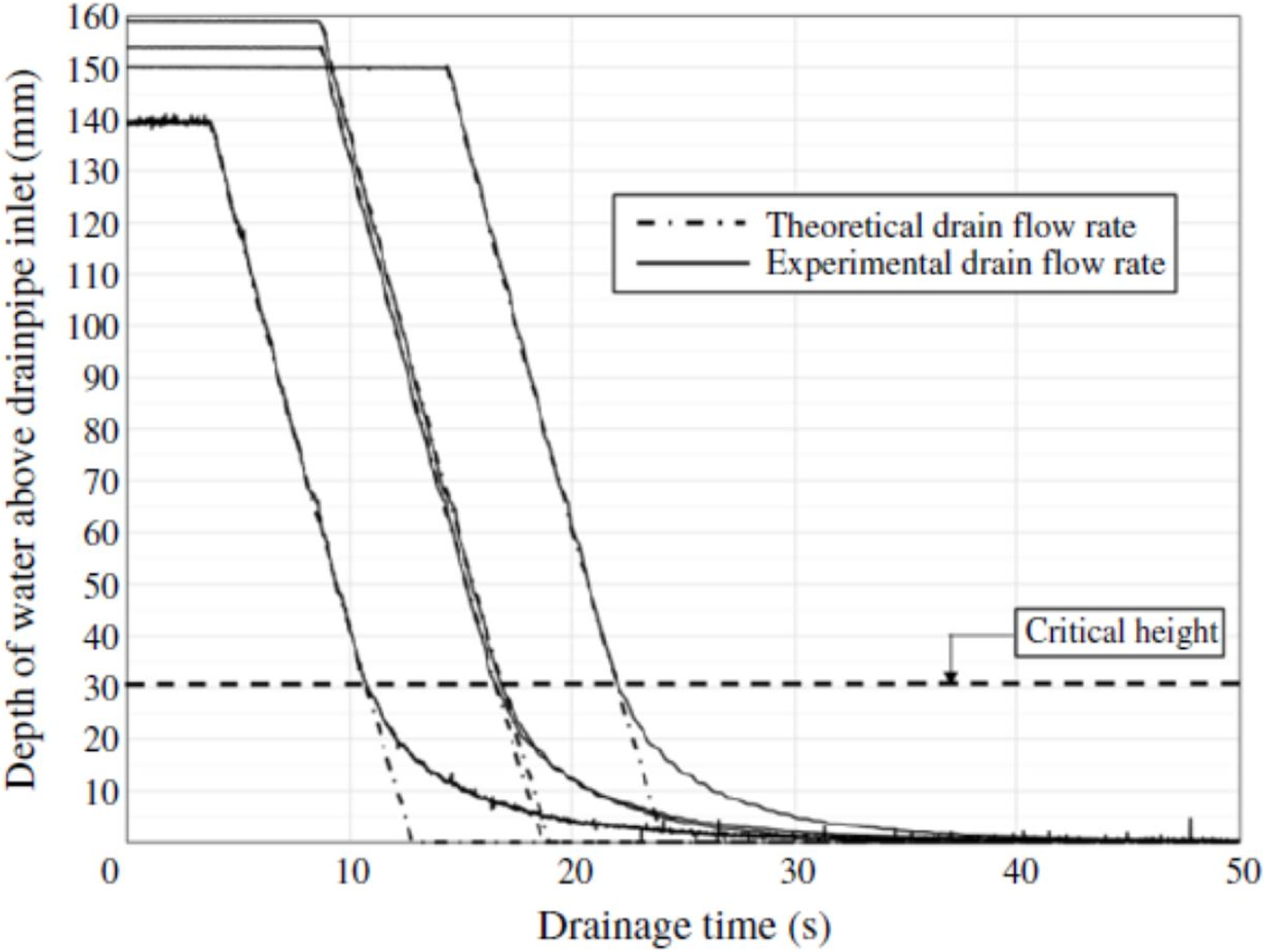}
\caption[]{Experimental and theoretical drain flow rate for a 49.8-mm drainpipe. In the figure, the $C_d$ (discharge coefficient) is set to 0.78.} \label{fig:Drainflowrate}
\end{figure}

Figure~\ref{fig:Drainflowrate} presents a comparison of the experimentally and theoretically determined water depth variation versus the drainage time for a 48.9-mm-diameter drainpipe. In this figure, four complete draining processes are presented. Because the horizontal cross-sectional area of the drain tank was independent of height, the draining flow rate could be estimated based on the temporal variation of the water depth (between the impulsive starting point and just before the formation of the plughole vortex). In the figure, there are four solid lines representing four experimental values of different initial water depths and four slanted dotted lines, mostly superimposed on the solid lines, representing the theoretical values corresponding to the respective experiments. There are three different regions for each draining case: a constant water depth, a linear slope, and an exponential decay. Before starting the draining of water, the constant water depth was at rest, and after the impulsive start of water draining, the outlet of the drainpipe let the water flow out, which caused a linear decrease in the water depth. Similar results were reported by \cite{Farrow&Hocking2006} for their 2-D viscous simulation and by \cite{Forbes&Hocking2007} for their 2-D inviscid model. In addition, Forbes \& Hocking considered two-dimensional, inviscid, and irrotational flow of a two-layer fluid in a tank. Because the lighter fluid was continuously recharged at the top of the tank during the complete draining of the heavier fluid, they observed the interface height at the centre of the tank as a function of time, which resulted in some discrepancies with the current study. However, interestingly, they also observed a linearly proportional range of the interface height, as well as a non-linear range.  \par

As Fig.~\ref{fig:Drainflowrate} shows, up to a certain depth, the change in the water level determined from the experiment agrees with the theoretically calculated value, and the two deviated close to the bottom surface. It should be noted that when the free surface of the water reached the point of deviation, the flow rate of the water decreased exponentially until the completion of the draining, whereas the theoretical calculation suggested that the flow rate remained almost unchanged. It is interesting to note that these four different points of deviation occur at a common water depth, which can be termed the critical height (indicated by the horizontal dashed line in the figure for the current configuration of the draining experiment). The critical height can best be obtained by comparing the experimental draining test results to the corresponding theoretical calculation results for a particular test case. \par

This figure suggests that the critical height is independent of the initial depth of the water above the inlet of the drainpipe, which supports the work of LS. However, it was observed that an unsteady condition, such as presence of turbulence, of water draining generated a premature plughole vortex, which seemed to induce a decrease in the drain flow rate at a depth greater than the critical height. In addition, it can be conjectured that the critical height would tend to be proportional to the strength of the unsteady condition (e.g., the turbulent intensity, rotationality, etc.). \par

\subsubsection{Measurement of critical height from flow visualisation}

\begin{figure}
\centering
\includegraphics[angle=0, trim=0 0 0 0, width=1.0\textwidth]{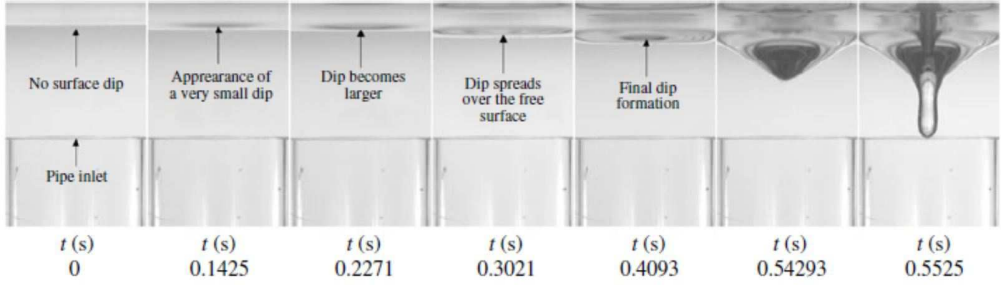}
\caption[]{Surface dip development} \label{fig:Initialsurfacedipdevelopment}
\end{figure}

The plughole vortex caused a noticeable surface dip to develop on the free surface of the water, and as the water depth decreased, the surface dip eventually entered the drainpipe (see Fig.~\ref{fig:Initialsurfacedipdevelopment}). It was observed that there was a certain time lag (0.5 to 0.6 s) between the start of the surface dip development and the moment just before the dip extended into the drain. It was also observed that the time lag was dependent upon the ratio of the free surface area of the tank to the cross-sectional area of the pipe. In the study on the surface dip development time by LS, it was observed that 'a dip forms above the drain on the surface of the bottom fluid and then grows so rapidly that it appears to the naked eye that the dip extends into the drain instantaneously'. Therefore, there seems to be a discrepancy between our results and those of the study by LS with respect to the surface dip development time. Although LS did not mention the exact time at which the dip entered the drain, it is believed that the condition that differed the most between the study by LS and the current study was the ratio of the free surface area of the tank to the cross-sectional area of the pipe (i.e., ratios of 81-5,184 in the study by LS and 152-10,610 in the present study). However, some recent numerical studies have confirmed that the dip development time depends on various fluid properties and container geometry. \cite{Zhou&Graebel1990} used a BIM scheme for their numerical simulation of axisymmetric draining and showed a transient surface profile that clearly illustrates the dip development time. \cite{Robinsonetal2010} and \cite{Forbes&Hocking2007} used different numerical simulation methods but, observed similar dip development time. \par

\subsubsection{Estimation of critical height from LS correlation}

To provide further supporting data regarding the critical height, the heights at which the surface dip was initiated were also estimated using the LS correlation for a water-air combination, and compared with the critical heights measured from the drain flow rate (A) and the images captured by high-speed camera (B) (see Table~\ref{tab:heightvariation}). The LS correlation can be expressed as follows: \par

\begin{equation} \label{eq7}
\frac{H_c}{a} = 0.69 \left( \frac{Q^2}{\left( 1 - \frac{\rho_2}{\rho_1} \right) ga^5} \right)^{\frac{1}{5}}
\end{equation}

\noindent where $H_c$ is the critical height, $a$ is the drainpipe radius, $Q$ is the average drain flow rate, $\rho_2$ and $\rho_1$ are the air and water densities, and $g$ is the acceleration due to gravity. Note that the critical height obtained from the LS correlation is estimated from the average drain flow rate based on the experimental data. In addition, the range of the flow rate was limited to the time just before the deviation of the drain flow rate began (see the horizontal dashed line in Fig.~\ref{fig:Drainflowrate}).

\begin{table} 
\begin{center}
~\caption{Critical height variation as a function of drainpipe radius. In the table, $H_c$ denotes the critical height, and the values shown were determined from the drain flow rate (A), the captured images (B), and the LS correlation (C).}
\scalebox{1.0}{
\begin{tabular}{ |c|c|c|c|c| }
\hline
 {\bf Drainpipe radius} & {\bf Average Q} & {\bf $H_c$, A} &  {\bf $H_c$, B} & {\bf $H_c$, C} \\
 mm 	& $\times 10^{-3} m^3/s$ 	&  mm 	&  mm 	&  mm \\ \hline
  6  	&   0.18920	&    9 	& 10.34	& 14.17 \\
  9.75	&   0.63097	&   14	& 16.55	& 22.95 \\
 13.1  	&   1.19390	&   18 	& 21.436& 29.60 \\
 15.95	&   1.84601	&   21	& 25.58	& 35.26 \\
 19.65 	&   2.77030	&   25 	& 28.435& 41.47 \\
 22		&   3.54073	&   28	& 32.43	& 46.11 \\ 
 24.9	&   4.62372	&   31	& 35.03	& 51.28 \\  
\hline
\end{tabular}} \label{tab:heightvariation}
\end{center}
\end{table}

Table~\ref{tab:heightvariation} shows the time-average drain flow rate over the linear range just before the formation of surface dip and a summary of the critical height variations determined by the different methods as a function of the drainpipe radius. There were some differences between the critical heights determined from processing of the captured images and the drain flow rate. In fact, the drain flow rate deviated from the theoretical value when the surface dip entered the drainpipe, i.e., at the time at which air entered the drainpipe and occupied some of the space inside. Therefore, because of the time lag between the start of the dip formation and the extension of the dip into the drain, the critical height calculated using the drain flow rate differed in some cases from the experimental value. The critical heights determined from the drain flow rate and the captured images also differed significantly from those determined from the LS correlation. Surprisingly, the critical height based on the drain flow rate was approximately 0.6 times that of the corresponding height based on the LS correlation. \cite{Robinsonetal2010} also observed some deviations from values estimated using the LS correlation in their VOF-based simulation of draining of two layers of fluid for a variety of fluid combinations. Although the measured critical height may include some measurement uncertainty, the critical height obtained in the LS experiment for a drainpipe with a 0.5-in (12.7-mm) radius, which was 0.82-in (20.828-mm), closely matches the critical height determined from image processing for the 13.1-mm-radius drainpipe used in this study. There may be two different reasons for these discrepancies. First, they used different fluid combinations, i.e., liquid-liquid and liquid-air combinations with various density differences. Second, there may be some scatter in the data used to develop the LS correlation, particularly for the water-air combinations. Therefore, there is likely to be some uncertainty associated with these three different methods in estimating the critical height. However, taking the above information into account, it can be demonstrated that the estimation of the critical height from captured images is one of the most precise measurements employed in this study. \par

\begin{figure}
\centering
\includegraphics[angle=0, trim=0 0 0 0, width=1.0\textwidth]{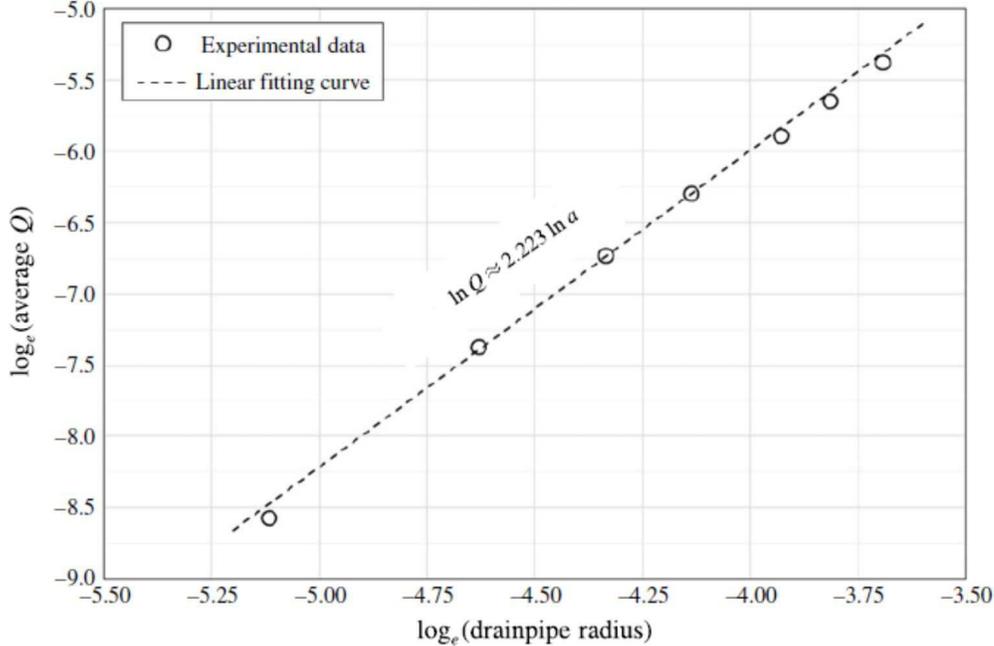}
\caption[]{Drain flow rate with drainpipe radius. In the figure, $Q$ is the volume rate of flow measured in $m^3/s$ and a is the drain pipe radius measured in $m$.} \label{fig:Drainflowratewithdrainpiperadius}
\end{figure}

Further examination of Table~\ref{tab:heightvariation} shows that the average volume flow rate increases with increasing drainpipe radius. Figure~\ref{fig:Drainflowratewithdrainpiperadius} is a log-log plot of the average drain flow rate versus the drainpipe radius. The plot shows that the average volume flow rate increases approximately quadratically with increasing radius of the drainpipe, as described by the relation $Q \approx ({\rm Constant}) a^{2.223}$, where $Q$ is the volume rate of flow measured in $m^3/s$ and a is the drainpipe radius measured in $m$. It was also observed that the critical heights was nearly linearly proportional to the drainpipe radius and the drain flow rate, even though the drainpipe lengths were all the same (see Table~\ref{tab:heightvariation}). It is important to note that the critical height of the water increases with increasing drain flow rate and that the drain flow rate increases with increasing drainpipe radius. \par

\subsection{Droplet-shaped bubble from plughole vortex}

\begin{figure}
\centering
\includegraphics[angle=0, trim=0 0 0 0, width=1.0\textwidth]{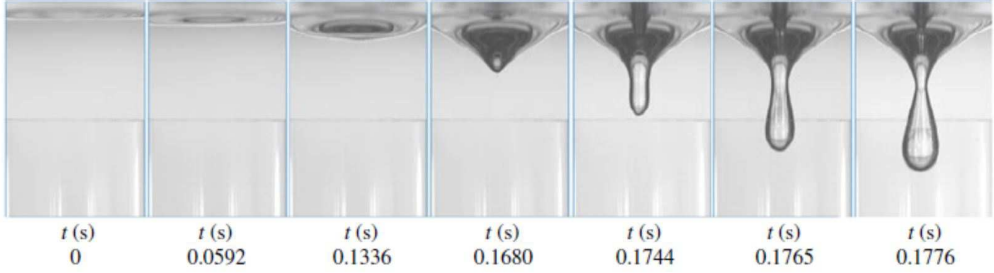}
\caption[]{Development of surface dip over a 31.9-mm diameter drainpipe.} \label{fig:Surfacedipdevt}
\end{figure}

The flow visualisation method used to observe the formation and gradual development of a plughole flow when the depth of water above the inlet of the drainpipe reached the critical height is explained in this section. In the early stage of dip development, the formation of a plughole vortex causes multiple concentric surface dips that spread over the free surface of the water, with a final dip extending into the inlet of the drainpipe within milliseconds (see Fig.~\ref{fig:Initialsurfacedipdevelopment}). Figure~\ref{fig:Surfacedipdevt} shows the surface dip development from the starting point of the final dip development to the complete bubble formation. The left-most image in the figure shows a slight dent on the water surface generating the dip development, and the images on the right show the extension of the final dip towards the inlet of the drainpipe. In the final stage of dip development, the surface dip takes the form of an inverted cylindrical column with a cone-shaped free surface. As the dip enters further into the drainpipe, the tip of the cylindrical column grows while the cone-shaped free surface remains unchanged. In addition, as the tip of the cylindrical column becomes larger, it forms a neck with a converging inlet zone, followed by a diverging outlet zone, and the neck diameter decreases with the expansion of the cylindrical dip. A bubble-forming instability at the tip of a bathtub vortex due to high rotation rates was also observed by \cite{Andersenetal2003}. In addition, they observed the formation of a neck and measured the decrease in the neck diameter with increasing bubble pinch-off time, which does not correspond to the work done in our study in terms of the bubble shape, size, or formation process. In this study, the expansion of the cylindrical dip was observed to result in a droplet-shaped air bubble attached to a cone-shaped free surface. It can be observed that, from the start of the final dip development, it took only 0.178 s to form the initial droplet-shaped bubble. In addition, the size of the air bubbles varied with the pipe diameter and the depth of the water above the inlet of the drainpipe. In general, the bubble size seemed to increase with increasing pipe diameter or decreasing water depth. \par

\begin{figure}
\centering
\includegraphics[angle=0, trim=0 0 0 0, width=1.0\textwidth]{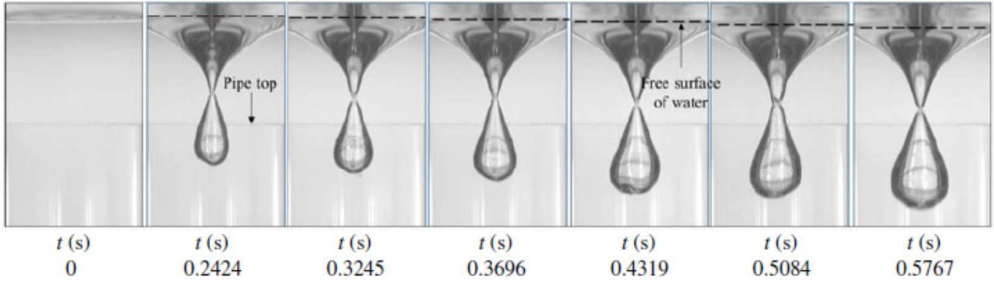}
\caption[]{Increase in bubble size with decreasing water depth for a 31.9-mm-diameter drainpipe.} \label{fig:Bubblesizevariation_39}
\end{figure}

Figure~\ref{fig:Bubblesizevariation_39} shows photographs of a droplet-shaped bubble immediately after detachment from the cone-shaped free surface at various water depths during a continuous draining process for a 31.9-mm-diameter drainpipe. During the draining process, the water depth gradually decreased with time, and thus it was necessary to select certain images to illustrate the variation in bubble size with respect to the water depth. The images were all captured using a high-speed camera at 7,500 fps during a complete draining process and only selective images showing the drainage time from the start of the final dip development are provided in the figure. \par

In the figure, the free surface of the water at different drainage times is marked with a dotted line. Interestingly, during the drainage process, the size of the bubble increased continuously, and this seemed to be caused by the decrease in the water depth. This might imply that a lower water depth lowers the static pressure around the droplet-shaped bubble, which allows the bubble size to increase. This idea will be explained and confirmed in due course through a numerical simulation. In addition, as the water depth decreases, the depth of the surface dip increases, and the location of the neck moves downstream.  \par

\begin{figure}
\centering
\includegraphics[angle=0, trim=0 0 0 0, width=1.0\textwidth]{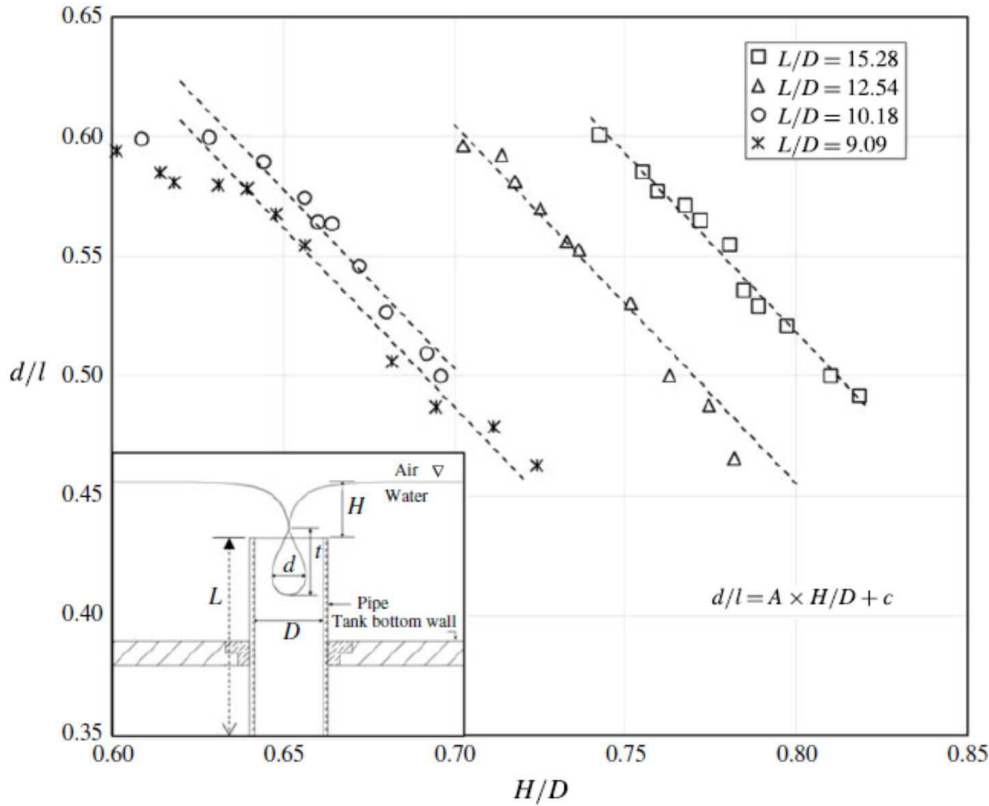}
\caption[]{Bubble size variation with change in water depth above the drainpipe.} \label{fig:Bubblesizevariationwchangingwaterdepth}
\end{figure}

Figure~\ref{fig:Bubblesizevariationwchangingwaterdepth} shows the variation in bubble aspect ratio with respect to the water depth above the inlet of the drainpipe for various ratios of drainpipe length to the diameter of the drainpipe ($L/D$). In the figure, the bubble diameter and depth of the water have been non-dimensionalised based on the bubble length and the diameter of the drainpipe, respectively. For a give $L/D$ ratio, the initial bubble diameter to length ($d/l$) ratio was at its smallest initially and increased as the drainage progresses. Interestingly, the figure shows that the $d/l$ ratio remained within a range of approximately 0.45 to 0.6 for all $L/D$ ratios considered. Moreover, the $d/l$ ratio seems to exhibit the linear trend expressed by equation ~\ref{eq8}. \par

\begin{equation} \label{eq8}
\frac{d}{l} = -1.5 \times \frac{H}{D} + C
\end{equation}

\noindent where $H$ is the depth of the water, which is less than the critical height, during the bubble formation. Interestingly, the slope of all of the linear relations was determined to be -1.5, and the constant $C$ (i.e., the $y$ intercept) ranged from 1.5 to 1.8 for the $L/D$ ratios mentioned, as shown in Fig.~\ref{fig:Bubblesizevariationwchangingwaterdepth}. The implication of this is that the $d/l$ ratio is linearly proportional to the $H/D$ ratio during most of the draining process.\par

Like the initial $d/l$ ratio for a drainpipe, the volume (i.e., size) of the bubble was also the smallest at the beginning of bubble formation, whereas its generation frequency was the highest compared to the rest of the draining process. The implication of this is that the bubble generation frequency decreases with the increasing bubble volume. In addition, because the bubble volume is also dependent upon the diameter of the drainpipe, a larger pipe diameter generates a larger bubble, and hence the bubble generation frequency for a larger-diameter drainpipe is also lower than for a smaller-diameter drainpipe. \par

\begin{figure}
\centering
\includegraphics[angle=0, trim=0 0 0 0, width=0.9\textwidth]{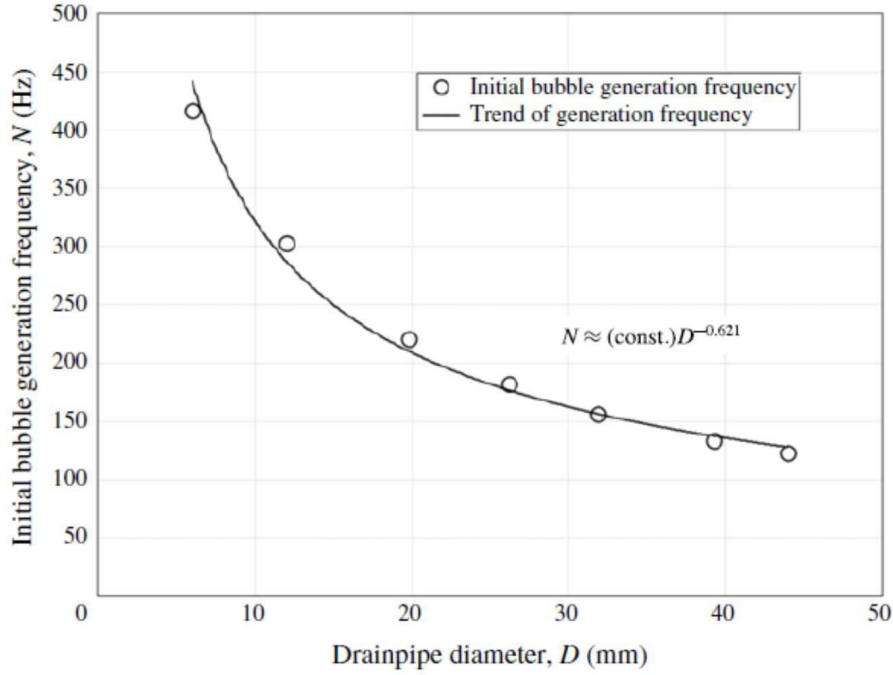}
\caption[]{Initial bubble generation frequency versus diameter of drainpipe.} \label{fig:Initialbubblegenerationfrequency}
\end{figure}

Figure~\ref{fig:Initialbubblegenerationfrequency} shows the variation in the initial bubble generation frequency with respect to the diameter of the drainpipe. As shown in the figure, as the diameter of the drainpipe increases, the decrease in the frequency of the bubble generation at the initial stage follows a power law relation (i.e., $N \approx ({\rm Constant}) D^{-0.621} $). In the case of a 6-mm-diameter drainpipe, the initial bubble frequency was observed to be as high as 410 Hz, whereas for a larger diameter of 44 mm, the frequency was only approximately 120 Hz. The important implication of this is that, as the pipe diameter increases, it takes more time to form larger-volume bubbles, which substantially reduces the frequency of bubble generation. \par

Interestingly, during the bubble formation process, intense noise generation was noted, particularly the sounds of instantaneous fizz and bubble sink draining, which seemed to result from the collapse of the bubble neck. The noise was measured to be approximately 105 dB at a distance of approximately 200 mm from the inlet for a 31.9-mm-diamater drainpipe. It was almost impossible to measure the velocity of the air at the bubble neck, which might be the source of sound that radiated from the collapse of the bubble neck. This point is addressed further in due course of this paper. \par

\subsection{Transformation of droplet-shaped air bubble to donut-shaped bubble ring}

\begin{figure}
\centering
\includegraphics[angle=0, trim=0 0 0 0, width=1.0\textwidth]{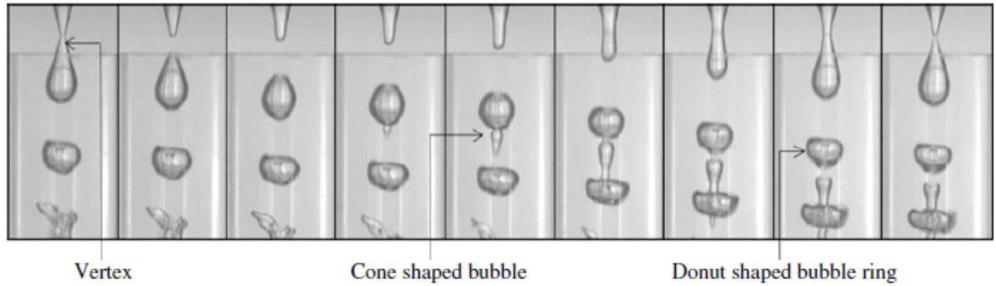}
\caption[]{Transformation of droplet into donut-shaped air bubble ring.} \label{fig:Droplet2donutshape}
\end{figure}

In the final stage of droplet-shaped bubble growth, the bubble neck collapsed, and the droplet-shaped air bubble detached from the surface dip due to the drag force of the surrounding flowing water. The detached droplet-shaped air bubble flowed with the drain water and was transformed from a droplet into donut-shaped bubble ring.  Photographs of the full process of air bubble transformation from a droplet-shaped air bubble to a donut-shaped bubble ring are shown in Fig.~\ref{fig:Droplet2donutshape}. In addition to the donut-shaped air bubble, the transformation process released another cone-shaped air bubble with a higher downward velocity than the donut-shaped bubble ring. \par

As shown in the figure, the full process of the transformation includes the formation of a typical droplet-shaped bubble, the generation of a concentric sphere, and finally, the formation of a donut shape. The transformation into a donut-shaped bubble indicates that, owing to the different velocity distributions in the cross section of the drainpipe, the flow velocity develops much more rapidly in the central region. In addition, the result of the analysis of captured images show that the vertex of the droplet-shaped bubble penetrated the bubble itself with a velocity approximately three to five times greater than the droplet-shaped bubble velocity yielding the donut-shaped bubble ring. In some cases, the cone-shaped air bubble also went through the same transformation and formed another small bubble. In most cases, the cone-shaped bubble penetrated or merged with its adjacent bubble ring, and flowed along with the draining water. The total transformation process took only 2 to 6 ms.

\subsection{Transformation of bubbly flow into annular flow in drainpipe}

\begin{figure}
\centering
\includegraphics[angle=0, trim=0 0 0 0, width=1.0\textwidth]{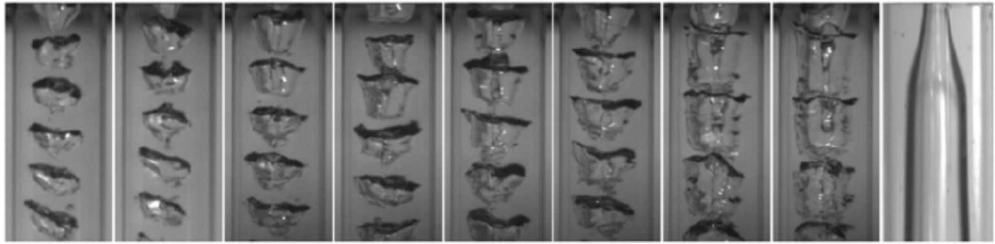}
\caption[]{Transition from bubbly flow to annular flow inside a drainpipe.} \label{fig:Bubbly2annularflow}
\end{figure}

At the start of bubble formation during a draining process, the initial donut-shaped bubble ring was relatively small, like the initial bubble volume, and increased with the decreasing water depth above the inlet of the drainpipe. As soon as the bubble ring formed, it flowed downwards with the draining water, even though it appeared to have a significant upward buoyancy force on the bubble. The buoyancy force was smaller than the drag force of the surrounding flowing water. It could be conjectured that because of the opposite directions of these forces, the shape of the bubble ring changed from circular cross-section to stretched and elongated. The larger bubble ring had a higher buoyancy force, which stretched the bubble ring more in the longitudinal direction. Figure~\ref{fig:Bubbly2annularflow} illustrates bubbly flow conditions inside a 6-mm-radius drainpipe at different stages of draining during a continuous draining process. The captured images shown in the figure were taken approximately 15 mm from the inlet of the drainpipe. It should be noted that the bubbles are more distorted in shape for the larger-diameter drainpipe. The images from left to right represent the initial to final stages of the draining process at different times during bubble formation. The shape and size of the bubble ring is also clearly visible. From the figure, it can be observed that initially smaller bubble rings caused bubbly flow in the drainpipe. As the droplet size increased with decreasing water depth, the bubble ring size also increased in the later stages. The increasing bubble size eventually caused slug flow in the drainpipe. However, at the end of the slug flow, each bubble ring merged with other rings, resulting in annular flow. As shown in the right-most image, the annular flow produced an air column that occupied most of the drainpipe volume. In the experiment, the flow rate in the drainpipe was at its lowest when it was annular flow. The draining process ended with annular flow in the drainpipe.

\section{Computational analysis and discussion}

The computation analysis described in this section was performed to explore the physics of the fluid flow phenomena. The purpose of our numerical work was to overcome the limitations of the plughole vortex experiment in obtaining the mean and fluctuating values of flow properties such as the pressure, velocity, and vorticity field, during the water draining. This section addresses the following subjects:
\begin{itemize}
\item Validation process: The validation of the computational methodology is primarily a prerequisite to show whether the calculation is accurate in terms of a comparison among a variety of drain flow rates for a 49.8 mm diameter drain pipe (Section 5.1).
\item Plughole vortex generation and bubble formation process: This subject pertains to the analysis and verification of a numerically generated plughole vortex and bubble formation during water draining, depending on the pipe configuration. In addition, the physics of bubble generation and growth in the drainpipe are explained (Sections 5.2 and 5.3).
\item Bubble transformation process: This subject pertains to the physics of bubble transformation from a droplet-shaped to a donut-shaped bubble ring, which was not observed in the experiment (Section 5.4).
\item Identification of vortex core region: This subject pertains to the distribution and temporal variation of primary vortices and vortex rings along the drainpipe, examined using a simple and efficient vortex core visualisation method (Section 5.5).
\end{itemize}
 
\subsection{Validation of the computational work}

\begin{figure}
\centering
\includegraphics[angle=0, trim=0 0 0 0, width=1.0\textwidth]{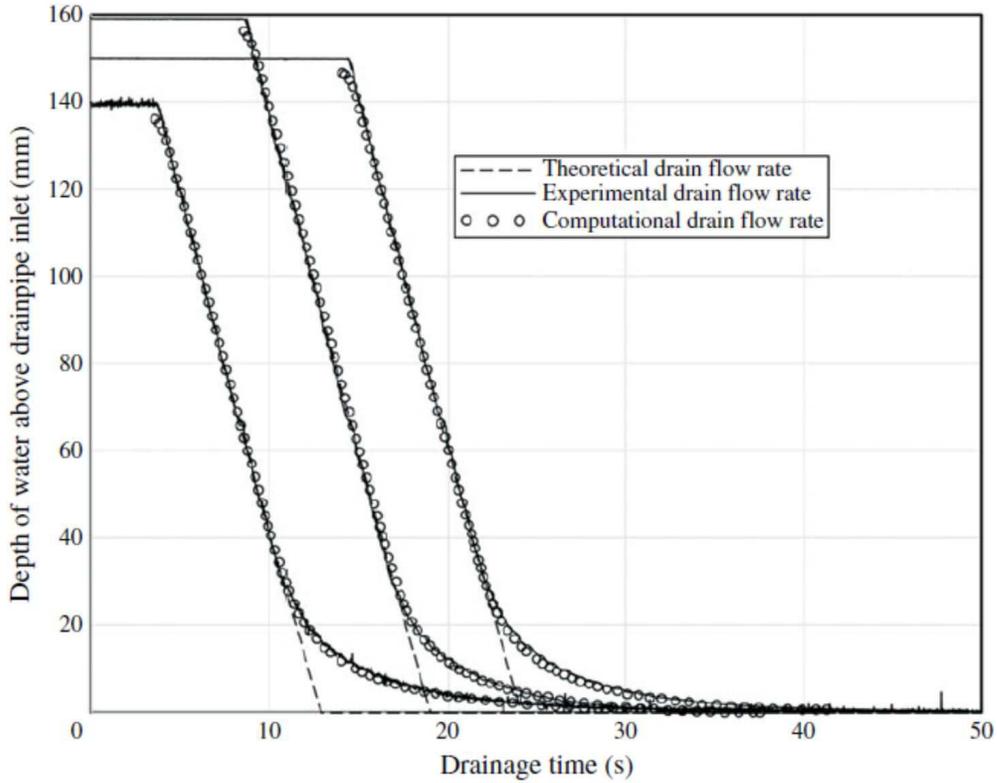}
\caption[]{Comparison of computational drain flow rate with experimental drain flow rate for a 49.8-mm-diameter drainpipe} \label{fig:ComprisonPipeFlow}
\end{figure}

\begin{figure}
\centering
\includegraphics[angle=0, trim=0 0 0 0, width=1.0\textwidth]{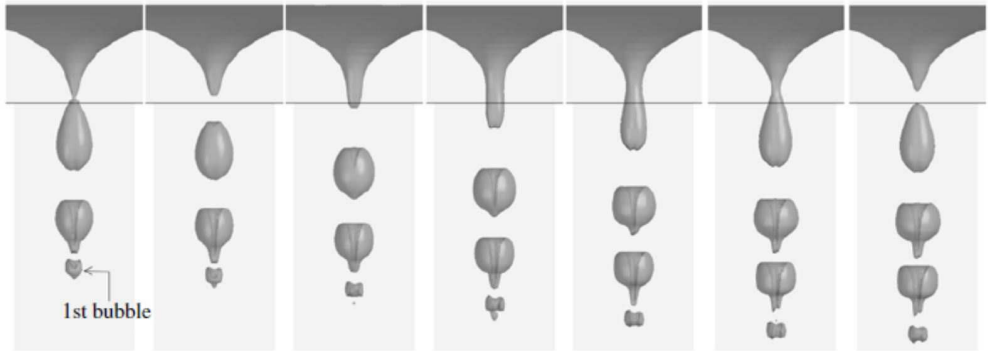}
\caption[]{Droplet-shaped bubble formation and transformation into a donut-shaped bubble ring}
\label{fig:NumDroplet2donut}
\end{figure}

A computational analysis was conducted for various cases by solving the Navier-Stokes equation (i.e., the realisable k-$\varepsilon$ turbulence model). Because such an experiment can be conducted in a limited environment, it would be appropriate to consider a variety of draining situations in the numerical simulation, allowing additional cases to be considered. Before going further, a validation of the computational methodology is necessary. Figure~\ref{fig:ComprisonPipeFlow} shows a comparison of the theoretical (eq. 4.1), experimental, and computational drain flow rates for a 49.8 mm diameter drain-pipe for the same tank dimension as of the experimental one. In computational analysis, the depth of the water at various stages of draining was obtained by tracking the air-water interfaces during the drainage time. As shown in the figure, it can be observed that the computational drain flow rate agreed surprisingly well with the experimental results. From the computational analysis, it was observed that the drain flow rate started to deviate from the theoretical drain flow rate when air bubble started to enter into the drain-pipe. The point of deviation for both the experimental and computational cases were also very close to each other. In addition, the bubble formation was also observed qualitatively (Fig.~\ref{fig:NumDroplet2donut}). Figure~\ref{fig:NumDroplet2donut} shows isometric views of an air–water interface representing a complete bubble formation. As shown in the figure, the formation of a droplet-shaped bubble and its transformation into a donut-shaped bubble ring can be clearly delineated. As described earlier, the simulation predicted the formation of a plughole vortex and its bubbly flow well, with the exception of the first droplet size, which was smaller than predicted. The numerical predictions of the transient bubble formation showed good agreement with the experimental observations (see Fig.~\ref{fig:Droplet2donutshape}). Therefore, the method could be used to describe the physics of this study. \par

\subsection{Free-surface shape of water}

\begin{figure}
\centering
\includegraphics[angle=0, trim=0 0 0 0, width=1.0\textwidth]{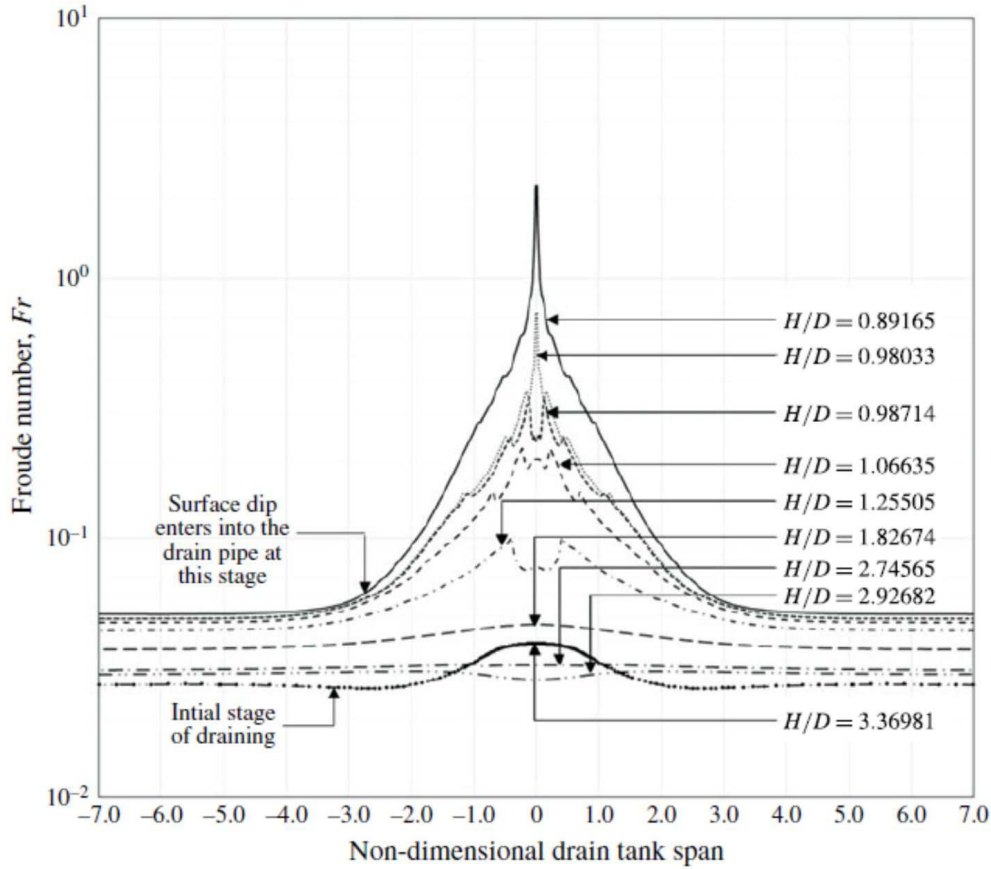}
\caption[]{Froude number at the free surface of the water just before bubble formation} \label{fig:FroudeNumberatthefreesurface} 
\end{figure}

In previous studies, plughole flow tests have been conducted to examine the free surface conditions during draining from a stationary tank both analytically and computationally. Most of such studies have mentioned the dependency on the Froude number, $Fr$, but they have also presented correlations with some other parameters, such as the surface tension (\citealt{Stokesetal2005}), the dimensionless drainpipe ratio (\citealt{Stokesetal2012}), and the fluid density ratio in the case of two layers of fluid (\citealt{Forbes&Hocking2010}).  In the present study, $Fr$ numbers were calculated for each case for the free surface of water during draining, and it was found that $Fr$ exceeds 1.0 momentarily when a surface dip enters the drainpipe. Similar observations were reported by \cite{Forbes&Hocking1995} and \cite{Zhou&Graebel1990}. In this study, $Fr$ numbers were calculated at the centre of the tank along the tank span, using the following equation: \par

\begin{equation} \label{eq9}
Fr = \frac{v_{down}}{\sqrt{gH_{local}}}
\end{equation}

\noindent where $v_{down}$ is the local downward velocity at the free surface of the water, $H_{local}$ is the local depth of the water above the inlet of the drainpipe and $g$ is the acceleration due to gravity. Figure~\ref{fig:FroudeNumberatthefreesurface} shows the calculated $Fr$ number along the tank span at various water depths during a continuous draining process. In the figure, the non-dimensional water depth $H/D$ is plotted against the non-dimensional tank span, where $H$ is the depth of the water far from the centre of the drainpipe, $D$ is the diameter of the drainpipe, and the non-dimensional tank span is the ratio of the tank width to the drainpipe radius. A bump-like rise in the bottom line at the centre of the tank indicates that the $Fr$ number increases abruptly because of the impulsive start to the draining, and as the draining progresses, the profiles became uniform throughout the tank span, which indicates stable draining. Interestingly, the $Fr$ number at the centre of the drain tank increases abruptly when the depth of water reaches a certain level. A fluctuation in the $Fr$ number above the centre of the drainpipe indicates surface rippling on the free surface of the water and a sharp increase in the distribution, which crosses the value 1.0, indicating a direct drawdown of the surface dip into the drainpipe. \par

\begin{figure}
\centering
\includegraphics[angle=0, trim=0 0 0 0, width=1.0\textwidth]{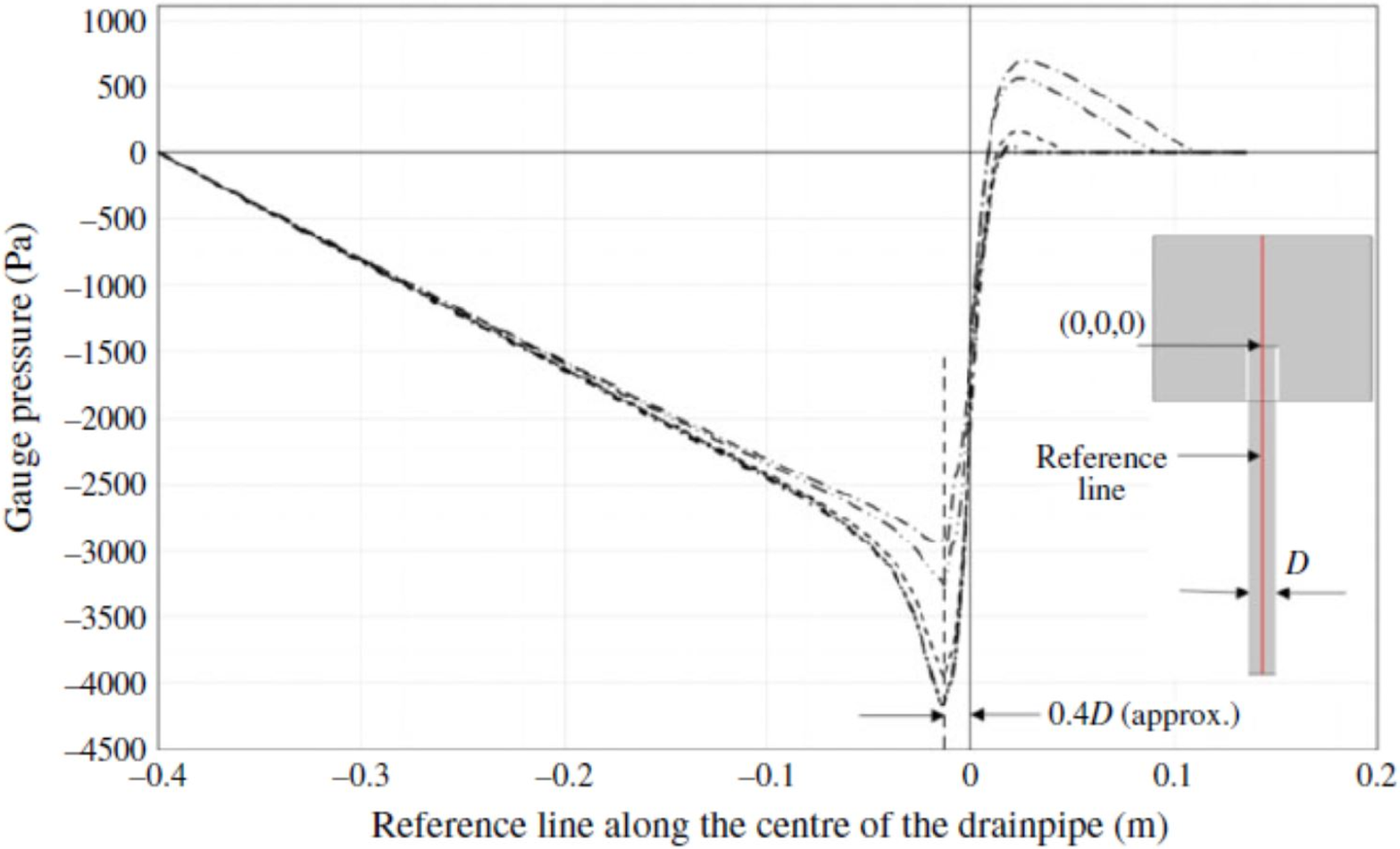}
\caption[]{Pressure variation along the centre line of the drainpipe} \label{fig:NumCentralpressurevariation}
\end{figure}

The main reason for the surface dip development along the inlet is still unclear. The low-pressure region under the water surface is one of the reasons that can explain the dip development over a free surface. To maintain zero pressure (i.e. atmospheric pressure) on a free surface, the pressure balance yielding a deformation of the free surface must be known. Figure~\ref{fig:NumCentralpressurevariation} shows the pressure variation along the reference line of the drainpipe. This line is located at the centre of the drainpipe, and the origin (0, 0, 0) was set at the centre of the inlet of the drainpipe. The horizontal and vertical axes represent the reference line and gauge pressure, respectively. As shown in Fig.~\ref{fig:NumCentralpressurevariation}, the pressure distribution can be divided into two regions--one above and the other below the inlet of the drainpipe. In the region above the inlet of the drainpipe, the pressure is at a maximum below the water surface. The minimum is in the region below the inlet of the drainpipe (i.e. at approximately 0.4 $D$ from the inlet of the drainpipe), and the pressure rises to atmospheric pressure. It was also observed that the negative pressure near the inlet of the drainpipe depends mainly on the vertical length of the drainpipe. Based on this fact, it can be determined that as the vertical length of the drainpipe increases, a decrease in the negative pressure (i.e., vacuum pressure) could occur close to the inlet of the drainpipe. Moreover, the vena contracta, which is a point in a fluid stream where the diameter of the stream is the least and fluid velocity is at its maximum, also plays a significant role in the decrease in the negative-pressure region. Interestingly, steady negative pressure is observed until bubble formation begins, whereas during bubble formation, the pressure varies substantially with time. \par

\begin{figure}
\centering
\includegraphics[angle=0, trim=0 0 0 0, width=1.0\textwidth]{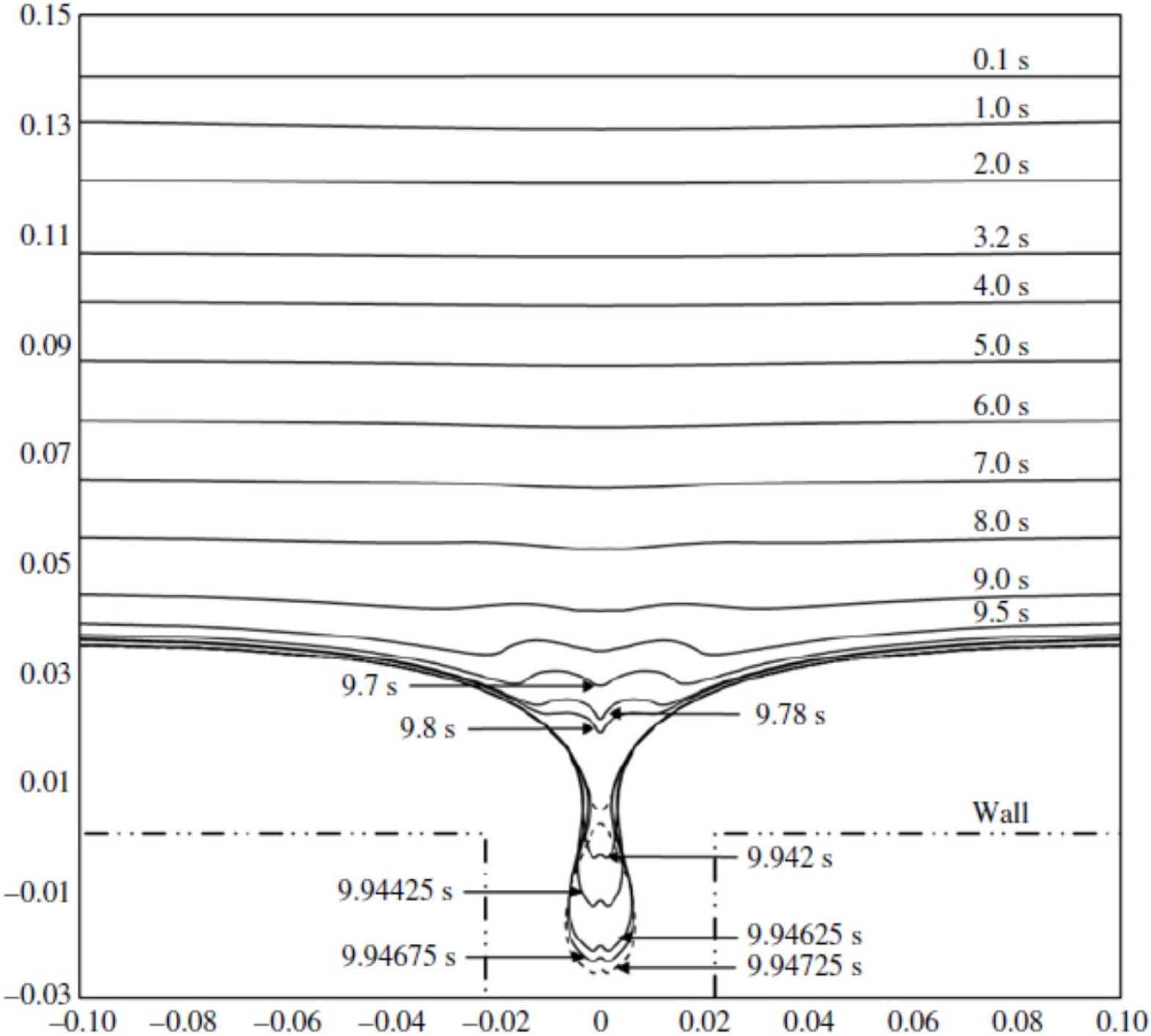}
\caption[]{Computational surface dip propagation into drainpipe, and formation of bubble droplet} \label{fig:NumSurfacedippropagation}
\end{figure}

Figure~\ref{fig:NumSurfacedippropagation} shows the free surface profile at different stages of a continuous draining. The figure shows that the shape of the free surface starts as a flat horizontal plane and gradually decreases in height. When it approaches the critical height (at approximately 9.6 s), it suddenly exhibits a surface rippling, which continues until the surface dip enters the drainpipe. These flow phenomena were also observed by \cite{Stokesetal2008}, who analysed the shape of a surface dip for various water depths and Froude numbers ($Fr$). They observed similar surface rippling, which they called a 'shelf effect' due to a sudden increase in $Fr$ from 0.008 to 0.16. In the present study, surface rippling also occurred at a Froude number between approximately 0.02 and 0.50. \cite{Zhou&Graebel1990} referred to this as 'surface oscillation', and attributed it to the 'draw-down provided by a sink being weak compared to the gravity force'. They also showed that a substantially large Froude number (much larger than 1.0) can ensure oscillation-free drainage, which is consistent with the result obtained in this study (see Fig.~\ref{fig:FroudeNumberatthefreesurface}). In addition, it was observed that as the free surface area inside the tank became smaller, the influence of a plughole dominated and thereby quickly dampened any existing oscillations. However, no such a surface rippling was observed during our experiment. In the computational analysis, the surface rippling did not have a significant effect on bubble formation, with the exception of the first few bubbles, which were smaller in size than those observed in the experiment, as shown earlier in Fig.~\ref{fig:Surfacedipdevt} (see also Fig.~\ref{fig:NumDroplet2donut}).

\subsection{Droplet-shaped bubble formation}

The surface dip development and the bubble formation process is also interesting. As shown earlier, as the water drains, a negative-pressure region close to the inlet of the drainpipe develops because of the flow of the water, which can be explained via the Bernoulli principle (i.e., following the law of energy conservation). Note also that the effect of the vena contracta should be taken into account in viscous flow. On the other hand, atmospheric pressure was maintained above the free surface of the water. Therefore, because the pressure balance must meet the requirement of stabilising the shape of the free surface above the centre of the drainpipe inlet, a surface dip is induced. Initially, an inverted-cone shape develops, as shown in Figs.~\ref{fig:Surfacedipdevt}, \ref{fig:NumDroplet2donut}, and \ref{fig:NumSurfacedippropagation}. As the draining progresses, the tip of the cone-shaped surface dip extends to the drainpipe inlet in the form of a cylindrical column. \cite{Zhou&Graebel1990} demonstrated a similar phenomenon of surface dip development during the draining of an axisymmetric cylindrical tank. However, they did not consider the effect of a draw-down pipe or the vena contracta in their simulation and hence did not observe the consequences of a surface dip extending into the drainpipe. As shown in Figs.~\ref{fig:NumDroplet2donut} and \ref{fig:NumSurfacedippropagation}, the tip of the cylindrical dip starts to expand when it enters the region of the drainpipe inlet, whereas the base of the cylindrical dip remains constant for a certain amount of time, forming a neck between the converging and diverging sections of the surface dip. Furthermore, as the tip of the surface dip expands further, it takes the form of a droplet-shaped bubble of air in the water. \par

\begin{figure}
\centering
\includegraphics[angle=0, trim=0 0 0 0, width=1.0\textwidth]{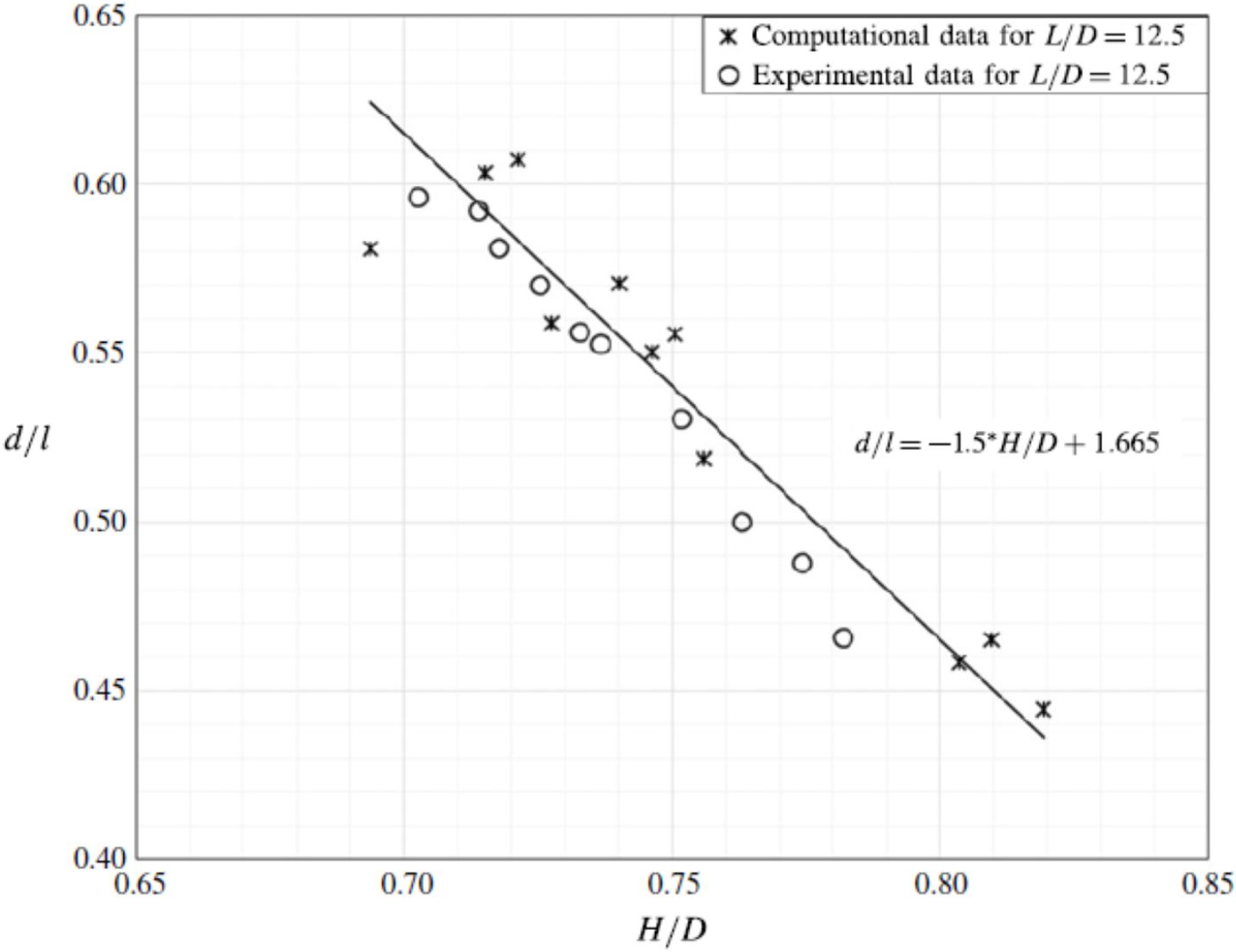}
\caption[]{Comparison of computational and experimental bubble sizes for $L/D$ = 12.5.} \label{fig:Comparisonofbubblesizes}
\end{figure}

Figure~\ref{fig:Comparisonofbubblesizes} shows a comparison of experimentally measured and numerically estimated bubble sizes. The bubble sizes are similar, and the range of the bubble diameter to length ratios ($d/l$) agrees well with the experimental result, as discussed earlier (see Fig.~\ref{fig:Bubblesizevariationwchangingwaterdepth}). The solid line in the figure, which lies between the experimental and computational results, shows that the numerical results also follow the linear trend given by the equation~\ref{eq8} for the experimental results. \par

\begin{figure}
\centering
\includegraphics[angle=0, trim=0 0 0 0, width=1.0\textwidth]{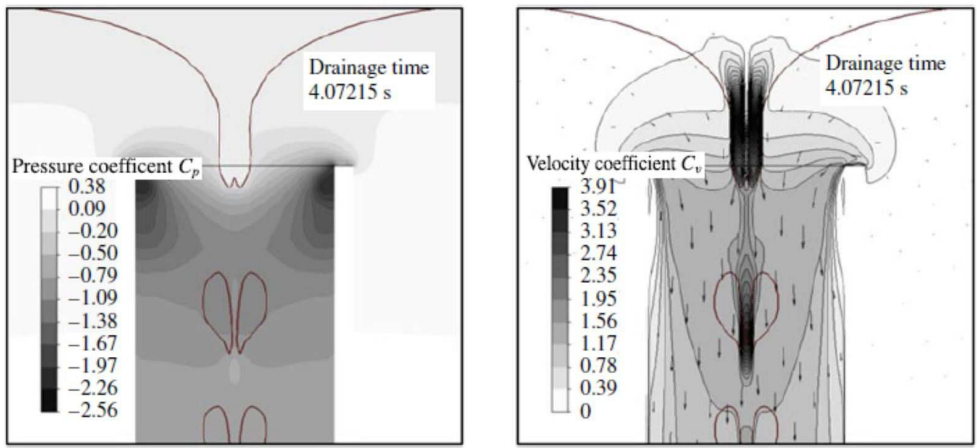}
\caption[]{Contours of pressure and velocity coefficient at the start of the droplet-shaped bubble formation.} \label{fig:NumPressureVelocitycoefficient}
\end{figure}

To better describe the pressure and velocity distribution during bubble formation, the contours of the pressure and velocity coefficients, $C_p$ and $C_v$, respectively around the centre region of a tank when a droplet-shaped bubble is formed are plotted in Fig.~\ref{fig:NumPressureVelocitycoefficient}. The pressure coefficient, $C_p$, was calculated from the ratio of the pressure difference from the reference pressure, $p_r$, to the dynamic pressure. In this study, for simplicity, the atmospheric pressure was taken to be the reference pressure. The dynamic pressure was defined as half of the product of the water density, $\rho_w$, and the square of the reference velocity, $v_{ref}$. The velocity coefficient, $C_v$, was calculated as the ratio of the velocity, $v$, to the reference velocity, $v_{ref}$, with the average outlet velocity taken to be the reference velocity. $C_p$ and $C_v$ can be expressed as follows: \par

\begin{equation} \label{eq10}
C_p = \frac{p-p_r}{0.5 \rho_w v^2_{ref}}; ~~~~ C_v = \frac{v}{v_{ref}}
\end{equation}

As Fig.~\ref{fig:NumPressureVelocitycoefficient} shows, the tip of the cylindrical surface dip starts to expand as it enters into the negative-$C_p$ region. There is a dimple at the tip of the cylindrical surface dip that closely resembles that reported by \cite{Forbes&Hocking2007}. The tip diameter expands as it enters deeper into the drainpipe, and the cylindrical surface dip takes the form of a droplet-shaped bubble. Because the droplet-shaped bubble has a low-pressure zone inside it, it tries to suck air from the atmosphere through the cylindrical neck. The $C_v$ contour shows that the air velocity at the centre of the neck is much higher than the adjacent water velocity. \par 

\begin{figure}
\centering
\includegraphics[angle=0, trim=0 0 0 0, width=1.0\textwidth]{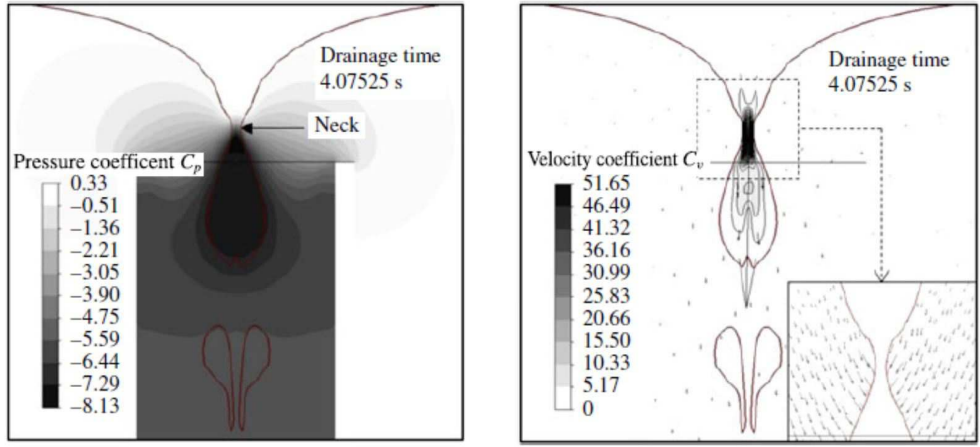}
\caption[]{Contours of pressure and velocity coefficient just before the separation of the droplet-shaped bubble.} \label{fig:NumPressureVelocitycoefficientbeforeseparation}
\end{figure}

As shown in Fig.~\ref{fig:NumPressureVelocitycoefficientbeforeseparation}, because the size (i.e., volume) of the droplet increases as it enters further into the drainpipe, the increasing size of the droplet seems to suck more air from the atmosphere through the neck, which induces a higher velocity, like a high-speed nozzle. Owing to the higher velocity of air in the neck passage, a high low-pressure zone is generated inside the neck, whereas a much higher pressure exists outside the neck. As a result, the negative pressure coefficient reached approximately -8.13 in this region. This pressure variation caused a strangling of the neck, which reduced the neck diameter dramatically and ultimately separated the droplet-shaped bubble from the surface dip, as a result of the downward dragging force of the surrounding flowing water. Figure~\ref{fig:NumPressureVelocitycoefficientbeforeseparation} clearly shows a strangling effect in the contour plot of $C_v$, as a result of the water enclosing the neck passage flowing towards the neck. However, the direction of flow was downward before the strangling effect occurred, as shown in Fig.~\ref{fig:NumPressureVelocitycoefficient}. \par

As described in Section 4.2, during the bubble formation process, an intense generation of noise occurred, particularly the sounds of instantaneous fizz and bubble sink draining, resulting from the collapse of the bubble neck. In the experiment, this phenomenon occurred so quickly that it was not possible to measure the air velocity at the neck, which might have been the source of the radiated sound. Therefore, based on the computational analysis, the velocity in the neck was analysed quantitatively. Figure~\ref{fig:NumPressureVelocitycoefficientbeforeseparation} indicates that as the bubble neck strangling occurred, the velocity coefficient of the air at the centre of the neck reached as high as 51.65 (i.e., over 100 $m/s$), which is reasonable, considering that the bubble volume can be formed within milliseconds. The time difference between the events shown in Figs.~\ref{fig:NumPressureVelocitycoefficient} and \ref{fig:NumPressureVelocitycoefficientbeforeseparation} indicates that the bubble volume expanded within only approximately 3 $ms$. The experimental and computational analyses of the bubble volume revealed similar tendency in terms of the temporal variation of the bubble volume. It should also be noted that close to the neck wall, the air velocity is the same as the surrounding water velocity at the interface (i.e., a no-slip condition exists), whereas the air velocity increases sharply, reaching the highest level at the centre of the neck. This implies that the neck of the droplet-shaped bubble acts like a nozzle. In fact, nozzle flow usually involves both high-velocity jets and turbulent eddies generated by shearing flow, which may be the source of radiated noise. Therefore, in this case, the high velocity of the air and the effect of the shear flow through the neck passage could be reasons for the radiated sounds of instantaneous draining noise (e.g. fizz and bubble sink draining) generated during bubble formation. \par

\subsection{Transformation from droplet-shaped air bubble into donut-shaped ring}

\begin{figure}
\centering
\includegraphics[angle=0, trim=0 0 0 0, width=1.0\textwidth]{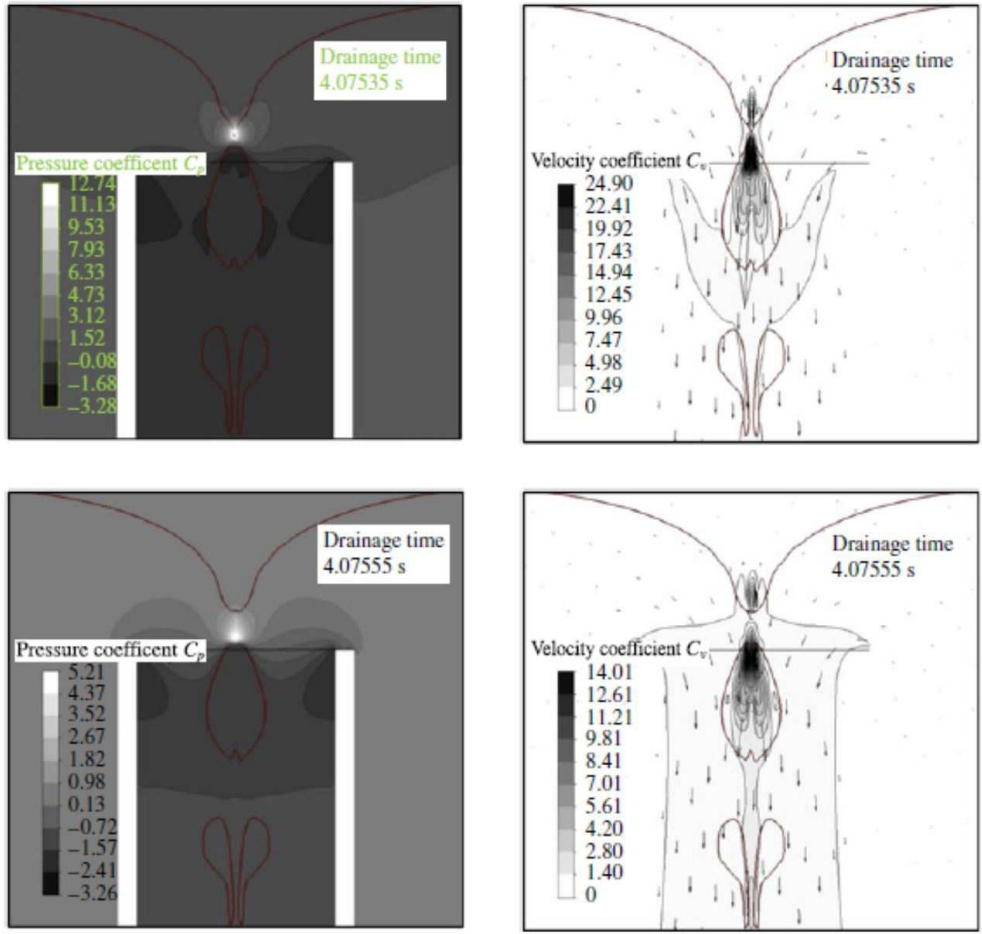}
\caption[]{Contours of pressure and velocity coefficients during bubble transformation.} \label{fig:NumPressureVelocitycoefficientduringtransformation}
\end{figure}

\begin{figure}
\centering
\includegraphics[angle=0, trim=0 0 0 0, width=1.0\textwidth]{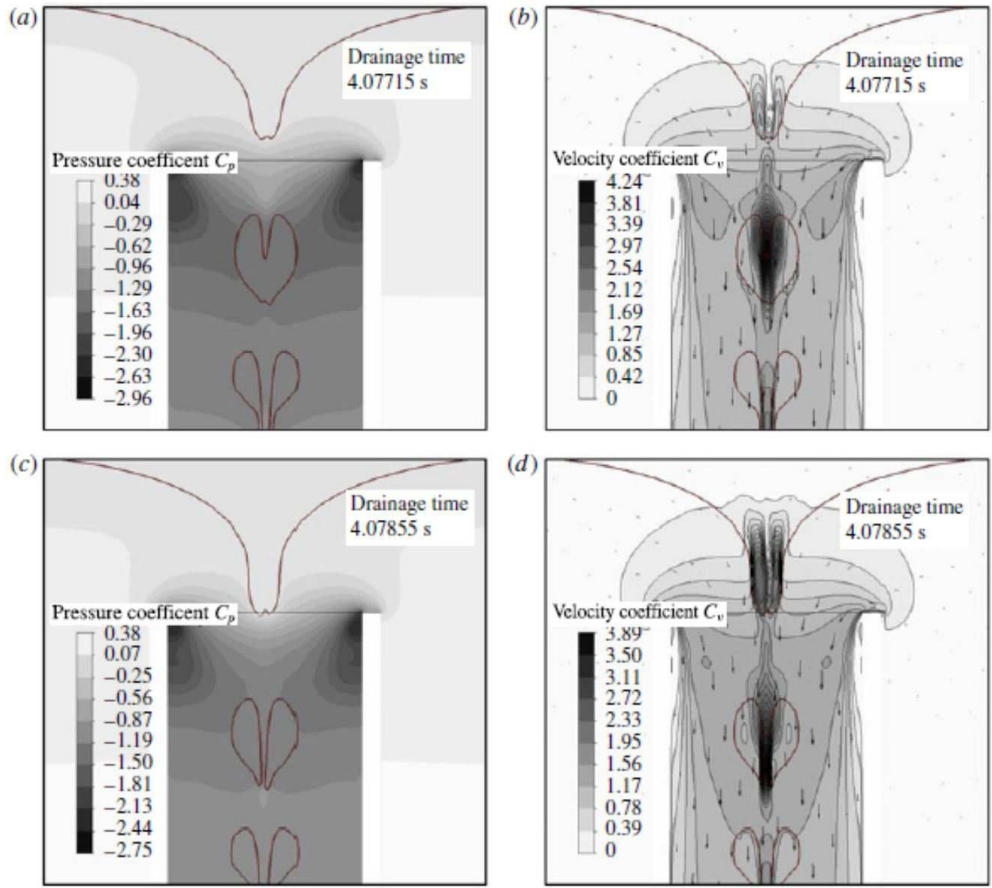}
\caption[]{Contours of pressure and velocity coefficients during bubble transformation.} \label{fig:NumPressureVelocitycoefficientduringtransformation2}
\end{figure}

To illustrate the process of transformation of a droplet-shaped bubble into a donut-shaped bubble ring, Figs.~\ref{fig:NumPressureVelocitycoefficientduringtransformation} and \ref{fig:NumPressureVelocitycoefficientduringtransformation2} show contours of the pressure and velocity coefficients at different drainage times during the bubble transformation. In Fig.~\ref{fig:NumPressureVelocitycoefficientduringtransformation}, the contour plot for the earlier drainage time (at 4.07535 s) shows that when the droplet-shaped bubble is separated from the surface dip, the impact and collision of the surrounding water moving towards the neck as a result of inertia result in a high-pressure zone (see the bright area in the pressure contours) in the merging neck. The simulation results also show that $C_p$ and $C_v$ reach nearly 12.7 and 24.90, respectively, in a very small region, within a fraction of a millisecond. This high-pressure and velocity inertia at the vertex of the droplet-shaped bubble and the vacuum pressure inside the droplet can be considered to pull the vertex inside the droplet at a much faster rate than the velocity of the droplet-shaped bubble. The induced vortices around bubbles would be another reason for this transformation. As shown in Fig.~\ref{fig:NumPressureVelocitycoefficientduringtransformation}, the $C_p$ and $C_v$ contours at a later drainage time (at 4.07555 s) illustrate a small dent formation around the vertex of the droplet-shaped bubble, as well as at locations where the velocity and pressure are significantly higher. \par

Figure~\ref{fig:NumPressureVelocitycoefficientduringtransformation2} shows the $C_p$ and $C_v$ contours during the penetration of the vertex into the droplet-shaped bubble. As the vertex of the droplet-shaped bubble enters the droplet, it looks like a slender water column penetrating the bubble and forming a bubble ring. The $C_p$ and $C_v$ contours of the figure show that, although the pressure inside the bubble was almost the same as the surrounding pressure, the penetration velocity was higher than the bubble velocity. In addition, because a cone-shaped bubble was clearly visible during the experiment, a similar shape (but not a clear shape) was also observed during the simulation. Although this might be true for a relatively coarser mesh compared to the size of a cone-shaped bubble, this problem can be resolved in future study by applying a finer mesh at the centre of the drainpipe and applying a shorter time step.\par 

\subsection{The size variation of droplet-shaped bubbles}

\begin{figure}
\centering
\includegraphics[angle=0, trim=0 0 0 0, width=1.0\textwidth]{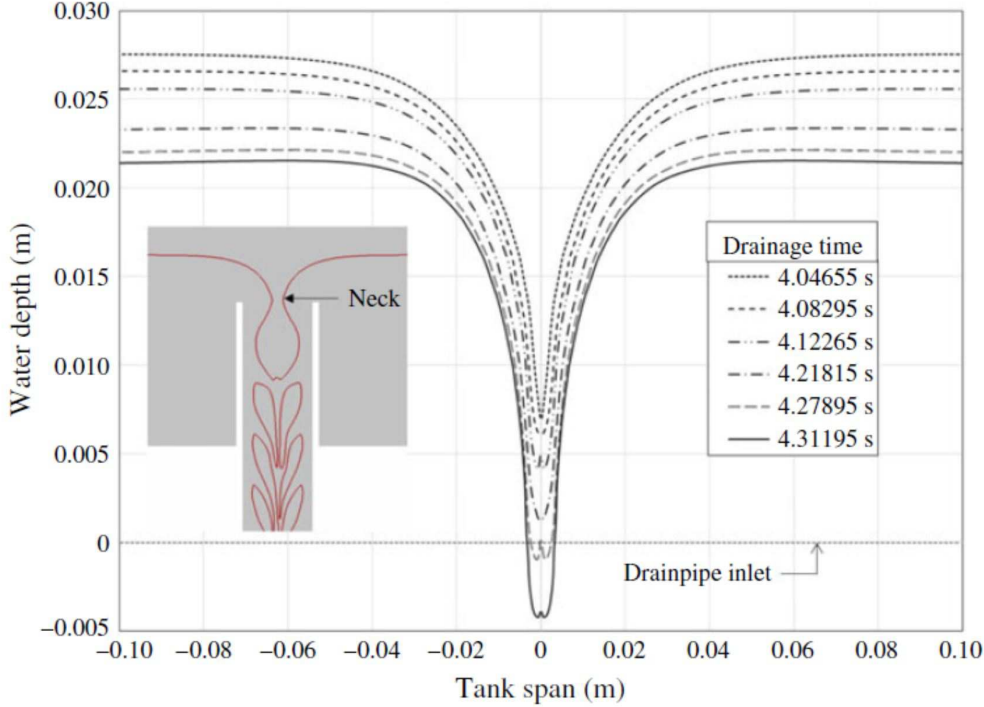}
\caption[]{Surface profile during bubble generation at various water depths.} \label{fig:Surfaceprofileduringbubblegeneration}
\end{figure}

The experimental observations showed that the size of a droplet increases with decreasing water depth above the inlet of the drainpipe, as shown earlier in Fig.~\ref{fig:Bubblesizevariation_39}. The computational observations revealed the same trend. Figure~\ref{fig:Surfaceprofileduringbubblegeneration} presents some fully developed surface dip profiles at various drainage times during bubble generation in a drainpipe. The surface profile shown here is for a free-surface dip just after the separation of a droplet-shaped air bubble. The top surface profile is for a drainage time of 4.04655 s, when the water depth was the highest compared to the others, and the tip diameter of the surface dip is small, compared to the rest of the profiles. As the drainage time increases and the water depth decreases, the tip of the cylindrical surface dip increases in diameter from the sharp peak present at an earlier stage. \par

The largest tip diameter was observed at a drainage time of 4.31195 s, when the water level was the lowest. As the depth of the water surface decreases, the static pressure in the water around the cylindrical dip gradually becomes lower, allowing the surface dip to expand, and then the cylindrical tip diameter increases. In a low-pressure region, the tip of the cylindrical dip expands and takes the form of a droplet-shaped bubble with a neck between the bubble and the free surface. The increased diameter of the cylindrical dip, i.e., the neck diameter, allows more air to enter the droplet-shaped bubble, and hence the droplet size increases with decreasing water depth. The experimental and computational observations indicate that at the later stages of draining, the bubble length increases significantly until annular flow eventually occurs in the drainpipe. \par

\subsection{Identification of vortex core region in a drainpipe}

During the computational analysis, the vortex generated by the draining was found to be irrotational because the Coriolis force was not taken into account during the simulation. Integration of the Coriolis force into the VOF simulation makes the simulation difficult and cumbersome and was thus avoided. During the experiment, the vortex was observed to be rotational, but the rotation was very weak and did not have much impact on bubble formation or its transformation. The effect of the Coriolis force on flow draining from a stationary tank, was examined in an experimental study by \cite{Shapiro1962}.\par 

To observe the vortex core inside a draining pipe, it is necessary to apply a vortex core visualisation criterion, such as the Helicity, Lambda criterion, Q-Criterion, or swirling strength. The swirling strength is an effective method for examining the shape of a vortex (see \citealt{Zhouetal1999}). The swirling strength is an imaginary part of the complex eigenvalues of a velocity gradient tensor. The velocity gradient tensor $\underline{D}$ can be expressed in terms of eigenvalues as follows:\par 

\begin{equation} \label{eq11}
\underline{D} \equiv \left[ v_r v_{cr} v_{ci} \right] 
\begin{bmatrix}
    \lambda_{r} & 0 & 0 \\
    0 & \lambda_{cr} & \lambda_{ci} \\
    0 & -\lambda_{ci} & \lambda_{cr} 
\end{bmatrix}
\left[ v_r v_{cr} v_{ci} \right]^{-1}
\end{equation}

\noindent where $\lambda_r$ is the real eigenvalue with a corresponding eigenvector, $v_r$, and $\lambda_{cr}\pm \lambda_{ci} i$ are the conjugate pair of the complex eigenvalues with complex eigenvectors, $v_{cr}\pm v_{ci}$. The strength of the local swirling motion can be quantified by $\lambda_{ci}$.\par

\begin{figure}
\centering
\includegraphics[angle=0, trim=0 0 0 0, width=1.0\textwidth]{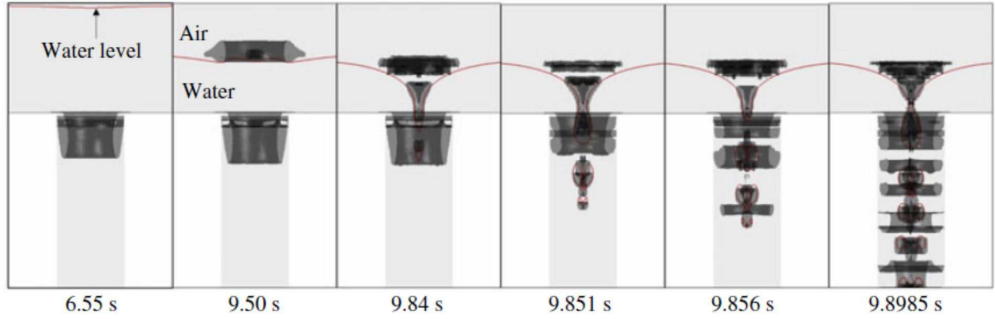}
\caption[]{Vortex ring observed in the simulation using swirling strength method.} \label{fig:Vortexring}
\end{figure}

The most notable feature of the swirling strength criterion is that it identifies not only the vortex core region but also the strength and local plane of the swirling. The swirling strength is positive if and only if the discriminant is positive, and its value represents the strength of the swirling motion around the local centre region. Figure~\ref{fig:Vortexring} delineates the vortex ring structure using the iso-surface for a fixed swirling motion ($\sim$ 8 $s^{-1}$) at various stages of draining. It was observed that the strength of the swirling motion varies over a wide range, from zero to greater than 20,000 $s^{-1}$, with the highest strength observed inside the bubble neck. However, the swirling strength of the majority of the pipe volume is between zero and 25 $s^{-1}$.\par

The computational analysis result reveal two distinct sources of vortices, one owing to the suction of air in the bubble, and the other from the vena contracta owing to the flowing water in the pipe inlet region. In Fig.~\ref{fig:Vortexring}, the left-most image, captured at a drainage time of 6.55 s, shows that, during normal draining without any surface dents, a steady vortex ring exists in the vena contracta region. The dent on the free surface of the water causes vortex rings to appear in the air. Multiple concentric vortex rings in the air can be observed at a drainage time of 9.84 s when the surface dip starts to enter the drainpipe. Images taken at drainage times of 9.851 and 9.856 s show that when a bubble separates from the surface dip it also separates a vortex ring from the vena contracta region, and the air bubble also contains a vortex ring inside it. The vortex ring generated in water from the vena contracta becomes larger in comparison to the vortex ring in the air bubble, and it flows downstream with the drain water, finally diminishing in the drainpipe wall. The right-most image in Fig.~\ref{fig:Vortexring} shows the vortex ring distribution during the bubble generation, which indicates that the vortex is strong in the vena contracta region and slowly diminishes along the downstream of the pipe. \par 

\section{Conclusions}
This study explored certain unique phenomena that occurs during the draining process, and attempted to explain the physics of a plughole vortex. Previous studies addressed how and when a surface dip develops and examined the effects of various fluid properties and conditions on plughole vortex. The current study was focused on describing in detail what happens when a surface dip from a plughole vortex enters a drainpipe. \par

In this study, the critical height was experimentally measured in two ways - using images captured with a high-speed camera and using measurement of the drain flow rate. The observed and calculated critical heights were compared, and the image-based analysis was found to be the most precise means of determining the critical height. A computational analysis was conducted to describe the physics of the experimental observations. The computational results showed good agreement with the experimental results, particularly for the formation of droplet-shaped bubbles and their transformation into a donut-shaped rings. A similar method for simulating a real sump with an outlet pipe in which a limited slippage exists between the gas and liquid phases was proposed by \cite{Robinsonetal2010} \par

The salient points of this study can be summarised as follows. First, the average volume flow rate increases approximately quadratically with increasing drainpipe radius (i.e., $Q \approx ({\rm Constant}) a^{2.223}$. The critical height is nearly linearly proportional to the drainpipe radius and drain flow rate. Second, a surface dip generates droplet-shaped bubbles when it enters a drainpipe. This seems to be result from the existence of a vacuum pressure zone at the inlet of the drainpipe due to the effect of the law of energy conservation and the vena contracta. Third, the ratio of the bubble diameter to the length ranges from 0.45 to 0.6, and is linearly proportional to the ratio of the water depth to the diameter of the drainpipe during bubble formation. Fourth, the initial bubble generation frequency follows a power law relationship (i.e., $N \approx ({\rm Constant}) D^{-0.621} $) with respect to the diameter of the drainpipe. Fifth, a droplet-shaped bubble is transformed into a donut-shaped bubble ring just after separation from the surface dip. The driving force for this transformation may be the high pressure generated at the vertex of the droplet-shaped bubble owing to its collapsing neck and inertia. Sixth, the initial bubble rings are comparatively smaller and create bubbly flow in the drainpipe. With decreasing water depth above the inlet of the drainpipe, the bubbly flow is transformed into slug flow and then annular flow in the drainpipe. Finally, the computational analysis results show that there are two distinct sources of vortices in a steady-state draining system: one is in the air due to bubble generation and the other is in the water due to the vena contracta. The vortices would have some effect on the droplet-shaped to donut-shaped bubble transformation. \par

The droplet-shaped bubbles observed in the computational analysis closely matched those observed in the experiment. Although the transformation of a droplet-shaped bubble into a donut-shaped ring was relatively smooth and noise free for pipes with diameter of less than 25 mm in both the computational and experimental observations, for larger-diameter pipes, the transformation was noisy and differed in the experiment from the computational observations. In our experiment, the bubble went askew just after separation from the surface dip, and the subsequent transformation distorted the bubble ring; in some cases, a bubble ring failed to even form. A possible reason for this distortion might be the existence of vacuum pressure inside the air bubble owing to the compressibility effect of air inside the neck of the bubble, which prevented air from entering. A computational analysis of incompressible fluids was carried out, but unfortunately the compressibility effect could not be calculated.\par

The pressure and velocity fluctuations at the bubble forming neck have not yet been verified experimentally, and if these could be verified, they could have a significant impact on some other applications. Surface dip development is due to the low pressure in the pipe suction area, and managing this low pressure would be useful in adjusting the critical height in cases of liquid draining or pump suction. \par

\vspace{0.5cm}

{\bf{Acknowledgement}}
This work was supported by "Human Resources Program in Energy Technology" of the Korea Institute of Energy Technology Evaluation and Planning (KETEP), granted financial resource from the Ministry of Trade, Industry \& Energy, Republic of Korea (No. 20164030201230). In addition, this work was supported by the National Research Foundation of Korea (NRF) grant funded by the Korea government (MSIP) (No. 2016R1A2B1013820).


\bibliographystyle{jfm}

\bibliography{JFM2017}

\end{document}